\documentclass[aps, prl, superscriptaddress,twocolumn,10pt]{revtex4-1}
\usepackage[left=1in, right=1in, top=1in, bottom=1in]{geometry}
\usepackage[toc,page]{appendix}

\usepackage[latin9]{inputenc}
\usepackage{bm}
\usepackage{tikz,pgfplots}
\usepackage{standalone}
\usepackage{braket}
\usepackage[retainorgcmds]{IEEEtrantools}
\usepackage{graphicx}
\usepackage{mathrsfs}
\usepackage{amsmath}
\usepackage{amssymb}
\usepackage{color}
\usepackage{amsfonts}
\usepackage{times,txfonts}
\usepackage{nicefrac}
\usepackage[colorlinks=true,linkcolor=blue,urlcolor=blue,citecolor=blue,pdfusetitle]{hyperref}
\usepackage{hyperref}
\usepackage[normalem]{ulem}
\usepackage{calrsfs}
\DeclareMathAlphabet{\mathcal}{OMS}{cmsy}{m}{n}
\usepackage{amssymb}
\usepackage{natbib}
\DeclareMathOperator{\tr}{tr}

\begin{document}
	
	\title{Reinforcement learning approach to non-equilibrium quantum thermodynamics}
	
	\author{Sofia Sgroi}
	\affiliation{Dipartimento di Fisica e Chimica - Emilio Segr\'e ,Universit\`a degli Studi di Palermo, via Archirafi 36, I-90123 Palermo, Italy}
\affiliation{	Centre for Theoretical Atomic, Molecular and Optical Physics, School of Mathematics and Physics, Queen's University Belfast, Belfast BT7 1NN, United Kingdom}
		\author{G. Massimo Palma}
		\affiliation{Dipartimento di Fisica e Chimica - Emilio Segr\'e ,Universit\`a degli Studi di Palermo, via Archirafi 36, I-90123 Palermo, Italy}
\affiliation{	NEST, Istituto Nanoscienze-CNR, Piazza S. Silvestro 12, 56127 Pisa, Italy}
			\author{Mauro Paternostro}
\affiliation{	Centre for Theoretical Atomic, Molecular and Optical Physics, School of Mathematics and Physics, Queen's University Belfast, Belfast BT7 1NN, United Kingdom}


	\date{\today{}}

\begin{abstract}
We use a reinforcement learning approach to reduce entropy production in a closed quantum system brought out of equilibrium. Our strategy makes use of an external control Hamiltonian and a policy gradient technique. Our approach bears no dependence on the quantitative tool chosen to characterize the degree of thermodynamic irreversibility induced by the dynamical process being considered, require little knowledge of the dynamics itself and does not need the tracking of the quantum state of the system during the evolution, thus embodying an experimentally non-demanding approach to the control of non-equilibrium quantum thermodynamics. We successfully apply our methods to the case of single- and two-particle systems subjected to time-dependent driving potentials.
\end{abstract}

\maketitle{}

	The design, development, and optimization of  quantum thermal cycles and engines is one of the most active and attention-catching research strand in the burgeoning field of quantum thermodynamics~\cite{art:rif.1,DeffnerCampbell,Marksman,Kosloff}. Besides being one of  the most important applications of thermodynamics, thermal engines play also a fundamental role in the development of the theory of classical thermodynamics itself. It is thus not surprising that the community working in the field that explores the interface between thermodynamics and quantum dynamics is very interested in devising techniques for the exploitation of quantum advantages for the sake of of realizing quantum cycles and machines~\cite{Marksman,Kosloff,Barontini2019}. The overarching goal is to operate at much smaller scales than classical motors and engines and enhance the performance of such devices so as to reach classically unachievable efficiencies~\cite{art:rif.1, art:rif.4, art:rif.10, art:rif.11}.

	However, the quasi-static approximation that allows us to describe thermodynamic transformations with an equilibrium theory does not hold for real finite-time thermal engines, which operate in non-equilibrium conditions. This is even more true for quantum engines: in order to exploit the potential benefits of quantum coherences, such devices should operate within the coherence time of the platforms used for their embodiment, which might be very short~\cite{Lindenfels2019,Peterson2019,Barontini2019}. Any finite-time process gives rise to a certain degree of  irreversibility, as quantified by entropy production, which enters directly into the thermodynamic efficiency of the process, limiting it~\cite{entropyproductionRMP}. Therefore, the control of non-equilibrium quantum processes is an important goal to achieve to enhance the efficiency of quantum engines~\cite{art:rif.13}. 
	
	For a closed system, a well-known quantum control approach consists of shortcuts-to-adiabaticity (STA)~\cite{TORRONTEGUI2013117, art:rif.16}. This approach has been successfully applied to various platforms~\cite{RevModPhys.91.045001, Palmero_2016, Mortensen_2018, PhysRevA.97.013628, REN201770, articlead, PhysRevLett.114.177206},
	and the possible application of STA to non-equilibrium thermodynamics has been explored~\cite{art:rif.13,art:rif.15, Dengeaar5909, PhysRevE.99.022110, PhysRevE.99.032108, PhysRevE.98.032121, unknown}. However, it bears considerable disadvantages as it requires extensive knowledge of the system dynamics. It is thus difficult to use STA as on-the-run experimental procedures. Moreover, they do not allow for the choice of the function characterizing the dissipative processes for the system  and it is currently very difficult to incorporate in a working STA protocol any constraint on the energetic cost of its implementation~\cite{Zheng2016,Santos2015}. Therefore, alternative approaches are necessary to improve our control power over quantum systems subjected to non-equilibrium processes.
	
	A possible approach to this problem is the use of machine learning techniques currently employed in growing number of problems. In particular, quantum physics is benefiting from machine learning in many ways in light of their capability to approximate high dimensional nonlinear functions that would be difficult to infer otherwise. Numerous applications have been developed, ranging from phase detection~\cite{mlph0, mlph1} to the simulation of stationary states of open quantum many-body systems~\cite{PhysRevB.99.214306}, from the research of novel quantum experiments~\cite{Melnikov1221} to quantum protocols design and state preparation~\cite{Porotti2019, PAPARELLE2020126266, PhysRevX.8.031086,Giordani2019,Giordani2020}, from the learning of states and operations~\cite{Innocenti2020,Harney2020, PhysRevX.8.031084} to the modelling and reconstruction of non-Markovian quantum processes~\cite{Melnikov1221,Banchi_2018}. In particular, problems of planning or control can be successfully addressed through reinforcement learning (RL)~\cite{book:rif.7}.
	
	In this Letter, we extend the range of quantum problems that can be tackled with machine learning approaches by demonstrating their successful use in the study of non-equilibrium thermodynamics of quantum processes. In particular, we propose an approach to reduce energy dissipation and irreversibility arising from a unitary work protocol using RL. Specifically, we employ a policy gradient technique to tackle out-of-equilibrium work-extraction protocols whose thermodynamic irreversibility we aim at reducing. Our methodology allows us to address this problem with only little knowledge of the system dynamics and to choose how to quantify dissipations. Our study provides a significant contribution to the development of control strategies tailored for physically relevant non-equilibrium quantum processes, thus complementing the scenario drawn so far and based on optimal control and STA.
	

\noindent
{\bf Background on reinforcement learning.--}
\begin{figure}
\includegraphics[width=0.6\columnwidth]{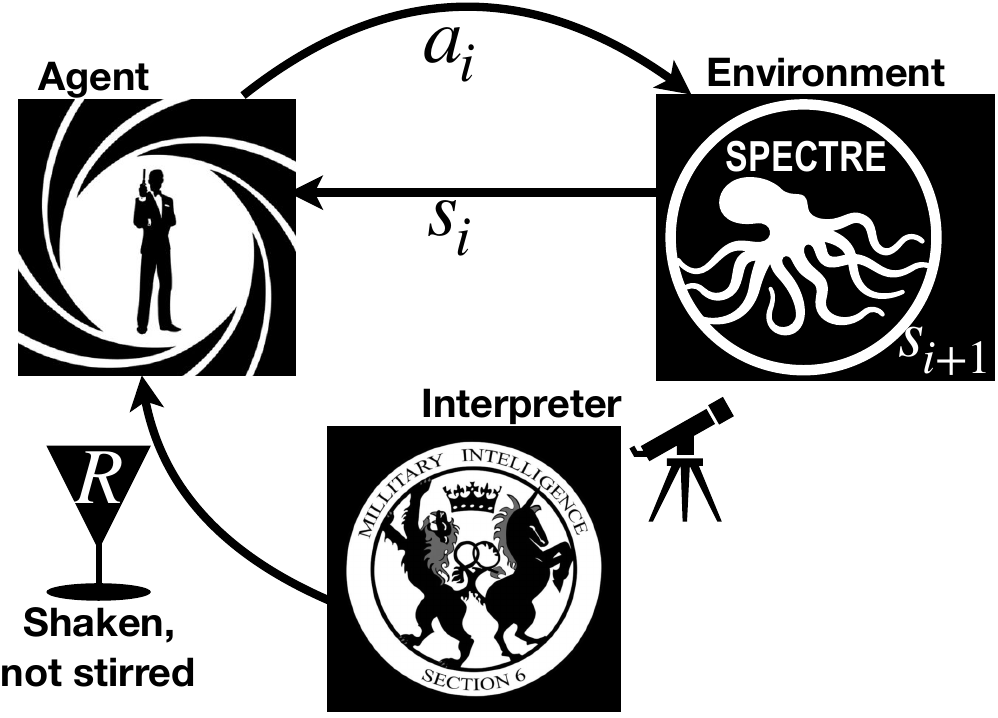}
\caption{Principles of RL: at the $i^\text{th}$ step of the protocol, an agent observes an environment, acquiring its state $s_i$, upon which he decides to implement action $a_i$. As a result, the state of the environment is updated to $s_{i+1}$. Based on the actions of the agent and the states of the environment, an interpreter decides to grant the agent  a reward $R$, which the agent aims to maximize.}
\label{007}
\end{figure}
	In the RL setting, an agent dynamically interacts with an environment and learns from such interaction how to behave in order to maximize a given reward functional~\cite{art:rif.21, book:rif.7}.
	The process is typically divided in discrete interaction steps: at each step $i$, the agent makes an observation of the environment state $s_i$ and -- based on the outcomes of their observations -- takes an action $a_i$. Based on this action, the environment state is updated to $s_{i+1}$ and we repeat the procedure for the new step. This is iterated for a given number of steps or until we reach a certain state, when a third party (an interpreter) provides the agent with a reward $R(s_0,a_0,s_1,a_1,...)$. Based on their past behaviour and the states of the environment, the agent change the way further actions are chosen so as to maximize the future reward (cf. Fig.~\ref{007}). This procedure is repeated for many epochs until, if possible, the agent learn how to reach the maximum reward.
	
	If the environment is completely observable, at each step the agent action and the reward depend only on the observation at the current step and the process is said to be a Markov decision process (MDP). In this case, we can describe the behaviour of the agent using a policy function $\pi(a_i|s_i)$. This represents the probability for the agent to choose the action $a_i$, given the state $s_i$ of the environment. In a policy gradient approach, we parametrize the policy function $\pi_{\theta}(a_i|s_i)$ with a set of parameters $\bm{\theta}$, and change them accordingly to the reward. This can be done using a gradient ascent algorithm. If the reward is given to the agent at the end of each epoch, as in our case, the gradient ascent reads \cite{web:ANN}
	\begin{equation}\label{gradient}
		\Delta\bm{\theta}=\eta R\sum_{a_i}\nabla_{\bm{\theta}}\log\pi_{\bm{\theta}}(a_i, s_i),
	\end{equation}
	where $\eta$ is the learning rate and the sum is calculated over the actions taken in any given {\it trajectory} $\{a_i\}_i$.
	
	For a continuous action space, we assume a certain shape for the policy function and use a function approximator for one or more parameters of the probability distribution \cite{book:rif.7}. Here we assume the policy function to be a Gaussian and take
	\begin{equation}\label{gauss}
	\pi_{\bm{\theta}}(a_i=a\vert s_i=s)=\exp\left[-\frac{(a-\mu_{\bm{\theta}}(s))^2}{2\sigma^2}\right]/({\sigma\sqrt{2\pi}}),
	\end{equation}
	where we treat $\sigma$ as an external parameter and use a neural network for the parametrization of $\mu$. 
	Based on our choice for $\pi_{\bm{\theta}}(a_i\vert s_i)$, the condition in Eq.~\eqref{gradient} is satisfied if the neural network is trained with a stochastic gradient descent method over the batch using the cost function
$		C=\frac{1}{2\sigma^2}\sum_{a_i}R|a_i-\mu_{\bm\theta}(s_i)|^2$.

\noindent
{\bf Physical system and methodology.-- }
	Let us consider a closed quantum system evolving under a time dependent Hamiltonian $H_S(t)$ within the time interval $[0,\tau]$. We want to control the system evolution using an additional Hamiltonian $H_\text{opt}(t)$ such that $H_\text{opt}(0)=H_\text{opt}(\tau)=0$.
	
	For simplicity, we consider $H_\text{opt}(t)=f_\text{opt}(t)M_\text{opt}$ where the operator $M_{opt}$ is kept fixed and we control the function $f_\text{opt}(t)$ (enforcing the boundary conditions $f_\text{opt}(0)=f_\text{opt}(\tau)=0$ so as to fulfil the requests made on the Hamiltonian) to optimize the process. The total Hamiltonian of the system during its evolution is thus
	\begin{equation}\label{control}
		H(t)=H_S(t)+f_\text{opt}(t)M_\text{opt}.
	\end{equation}
	We divide the system evolution in a certain number of discrete time steps. At each step $t_i$, the agent makes an observation $s_i$ and chooses an action $a_i$. This is done by extracting a random number according to Eq.~\eqref{gauss}, based on the prediction of the neural network for $\mu_{\bm\theta}(s_i)$. 
	We then take $f_\text{opt}(t)=a_i$ in the interval $[t_i, t_{i+1}[$. We limit the maximum and the minimum output of the network $|\mu_{\bm\theta}(s_i)|<\mu^*$ so that we can control the maximum amount of energy spent for the optimization. This is important when dealing with thermal engines, as we want to spend les energy for the control than what we extract from the process. This is done in parallel for a batch of systems and, at the end of the evolution, the neural network is trained on this batch and the corrisponding rewards. The procedure is repeated for many epochs, each time resetting the system and the Hamiltonian to the original state and value. The process is run again and the value of $f_\text{opt}(t)$ maximizing the reward over the batch is chosen.
	
	We now comment on the quantifier of irreversibility addressed in our study and the different approaches that we will consider to reduce the system dissipations. The first approach aims to reduce the mean entropy production of the system~\cite{art:rif.2, art:rif.3, art:rif.5, art:rif.6, art:rif.14, art:rif.1}, calculated as the relative entropy between the final state of the system $\rho(\tau)$ and the corresponding instantaneous thermal equilibrium state $\rho^{eq}(t)=e^{-\beta H_S(t)}/Z_S(t)$ with $Z_S(t)=\tr[e^{-\beta H_S(t)}]$ the partition function of the system. We thus consider 
	\begin{equation}\label{rel_entr}
	\Sigma=S(\rho(\tau)||\rho^{eq}(\tau))
	\end{equation}
	where $S(\sigma||\chi)=\tr[\sigma(\log\sigma-\log\chi)]$ is the quantum relative entropy~\cite{Vedral2002}.
	For this purpose, we use a Dense-layers Neural Network~\cite{book:rif.9} taking as inputs the time step and the density matrix. In this case, the agent reward is
$R=-\Sigma$,
which suits our goal well: the agent is rewarded by reducing the degree of irreversibility of the process.

	In the second approach, we assume to measure the energy of the system before the evolution \cite{art:rif.7, art:rif.8}. We consider the  case of non-degenerate energy levels and use as reward the square-root of the fidelity between the final state of the system and the corresponding adiabatic final state, thus having
$	R=|\langle \phi(t)|\phi_{ad}(t) \rangle|$.
	This approach too benefits of the use of a Dense-layers Neural Network with inputs embodied by the time step and the (pure) quantum state of the system.
		
		The third approach uses the same ideas laid out above. However, this time we want the model to be useful as a control technique even when we are not able to simulate or track the dynamics of the system. We thus use a different input, while still considering MDP. We use a Long-Short-Term-Memory (LSTM) Neural Network \cite{web:ANN} taking as inputs the energy measured at the beginning of the evolution, and the time steps.
		
		If the observation of our agent at a given time step contains all the informations about the initial state of the system and the control term of the Hamiltonian at any previous time, the knowledge of the current quantum state is no longer required in order to have a MDP. However, we can avoid to use such a large imput at each step if we use a LSTM network instead. The output of a LSTM network does not only depend on the input at a given time step, but also on all the previous imputs and outputs.
		Such neural networks can retain long-term dependencies and are widely used for tasks that involve sequential data, such as speech recognition.
		
		For these reasons, we just need to take measurements at the beginning and at the end of the evolution. Needless to say, this embodies a significant reduction on the practical complexity of the control protocol, as the scheme only requires two measurements, and thus leaves room for a non-demanding experimental implementations that does not need to track the evolution of the system.
		
	For inaccessible initial states of the system, or should one want to avoid performing a measurement at the start of the dynamics, if we assume the initial density matrix of the system to be always the same, we can still use a LSTM network in a way similar to the first approach, with just the time steps as inputs and a reward $R=-\Sigma$.
	 The advantage of our approach with respect to other techniques lies on the number of runs needed to achieve good performances, which is significantly smaller than what is needed from standard numerical optimization (see Ref.~\cite{SM} for details).
	
\noindent
{\bf Case studies.-- }
		We now apply our methods to simple yet physically relevant models. We address the cases of a single spin-like system exposed to a time-dependent field and a two-spin model with a time-dependent coupling. \\
{\it Single spin-$1/2$ particle in a time-dependent field}: Let us consider a qubit evolving in the interval $t\in[0,\tau]$ under an Hamiltonian (we take units such that $\hbar=1$)
	\begin{equation}\label{1spin}
		H_S(t)=\left[\sigma_x B_x(t)+\sigma_zB_z(t)\right]/2
	\end{equation}
	with $B^2_x(t)+B^2_z(t)=B^2_0$ $\forall t\in[0,\tau]$, thus modelling a spin subjected to a rotating magnetic field.
	We assume that the system is initialized in a thermal state at inverse temperature $\beta$. This is relevant only when we do not take an energy measurement at the beginning of the evolution. Our optimization Hamiltonian is $H_\text{opt}(t)=-f_{opt}(t)\sigma_y$ so that $H(t)=H_S(t)-f_{opt}(t)\sigma_y$.

	\begin{figure}[t!]
{\bf (a)}\hskip3.8cm{\bf (b)}\\
	\includegraphics[width=0.5\columnwidth]{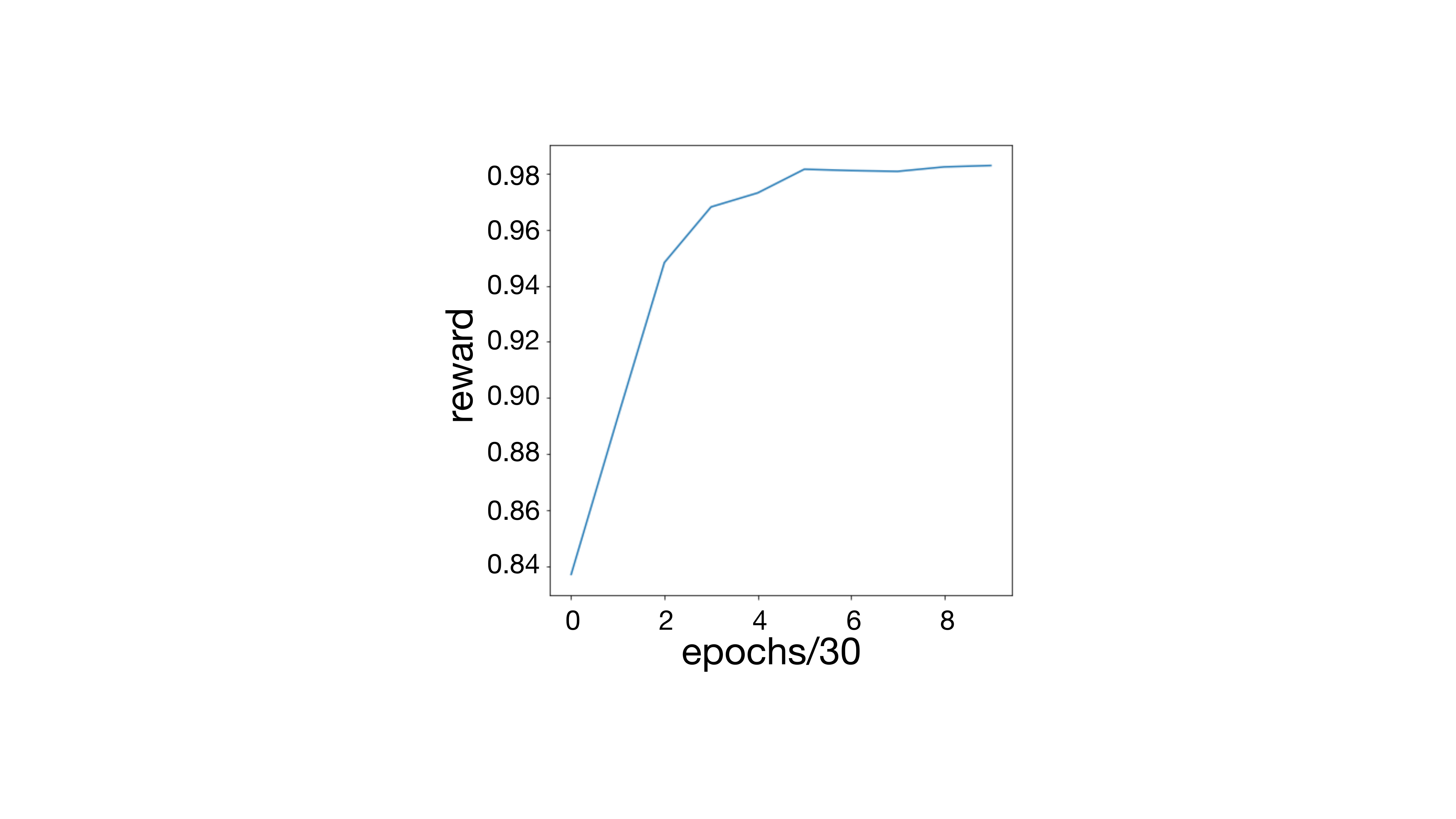}~\includegraphics[width=0.5\columnwidth]{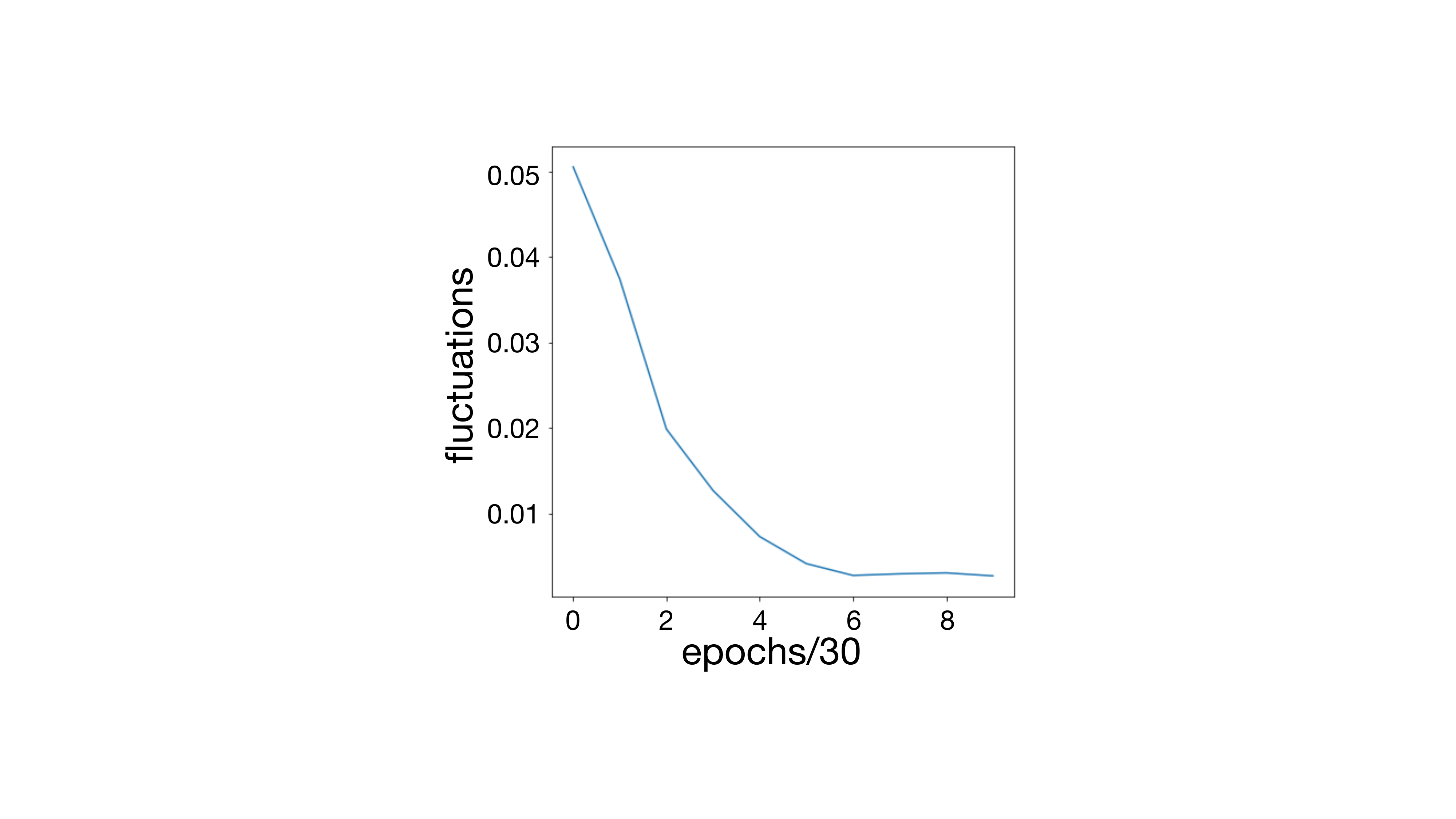}
		\caption{
			Panel {\bf (a)}: average reward over the batch and $30$ epochs as a function of the number of epochs of training. Panel {\bf (b)}: fluctuations around the average reward. The asymptotic behavior of both curves demonstrates success of the training.}\label{Figura:learning}
	\end{figure}
	\begin{figure}[b!]
	{\bf (a)}\hskip3.8cm{\bf (b)}\\
			\includegraphics[width=0.5\columnwidth]{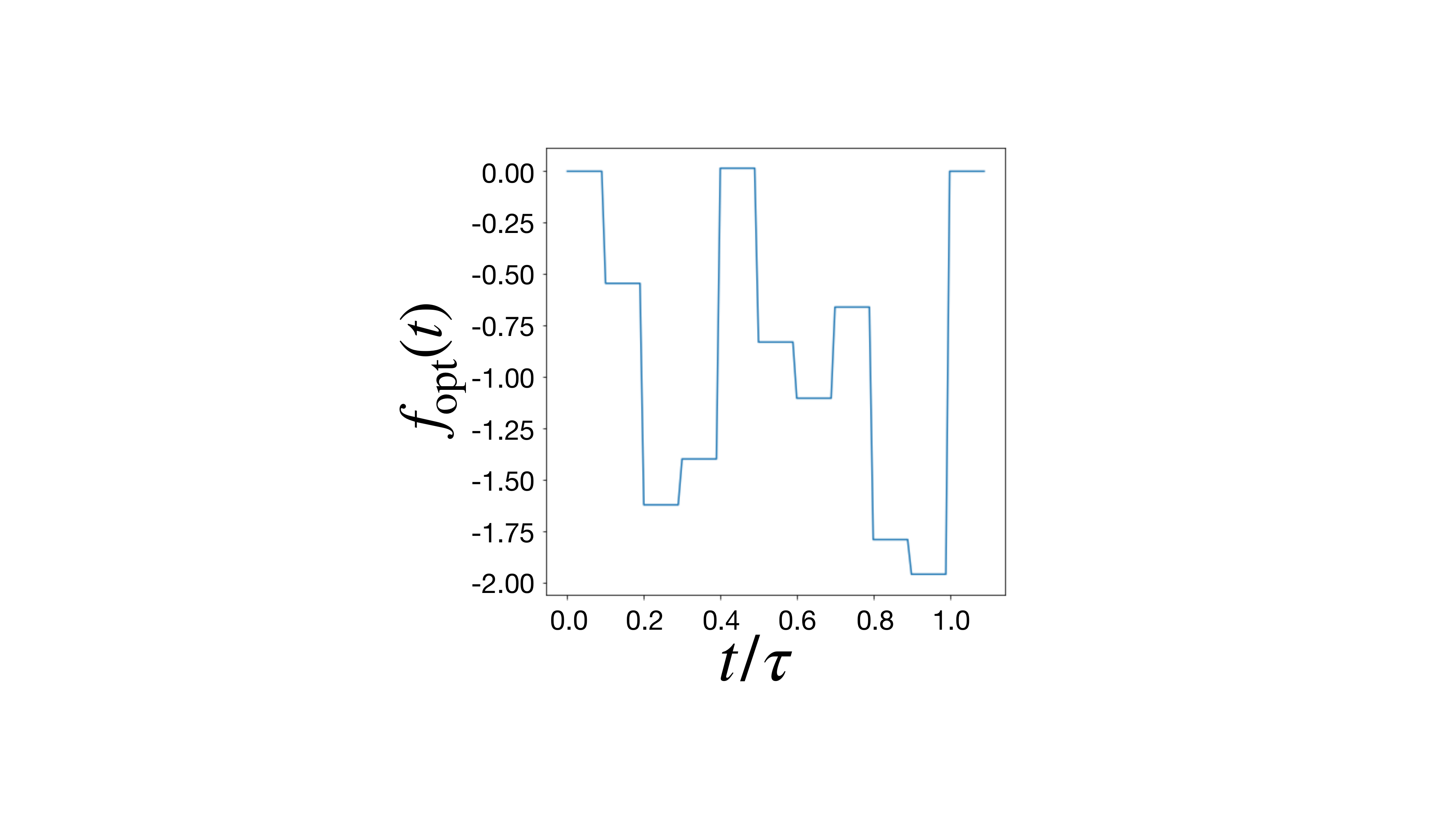}~\includegraphics[width=0.5\columnwidth]{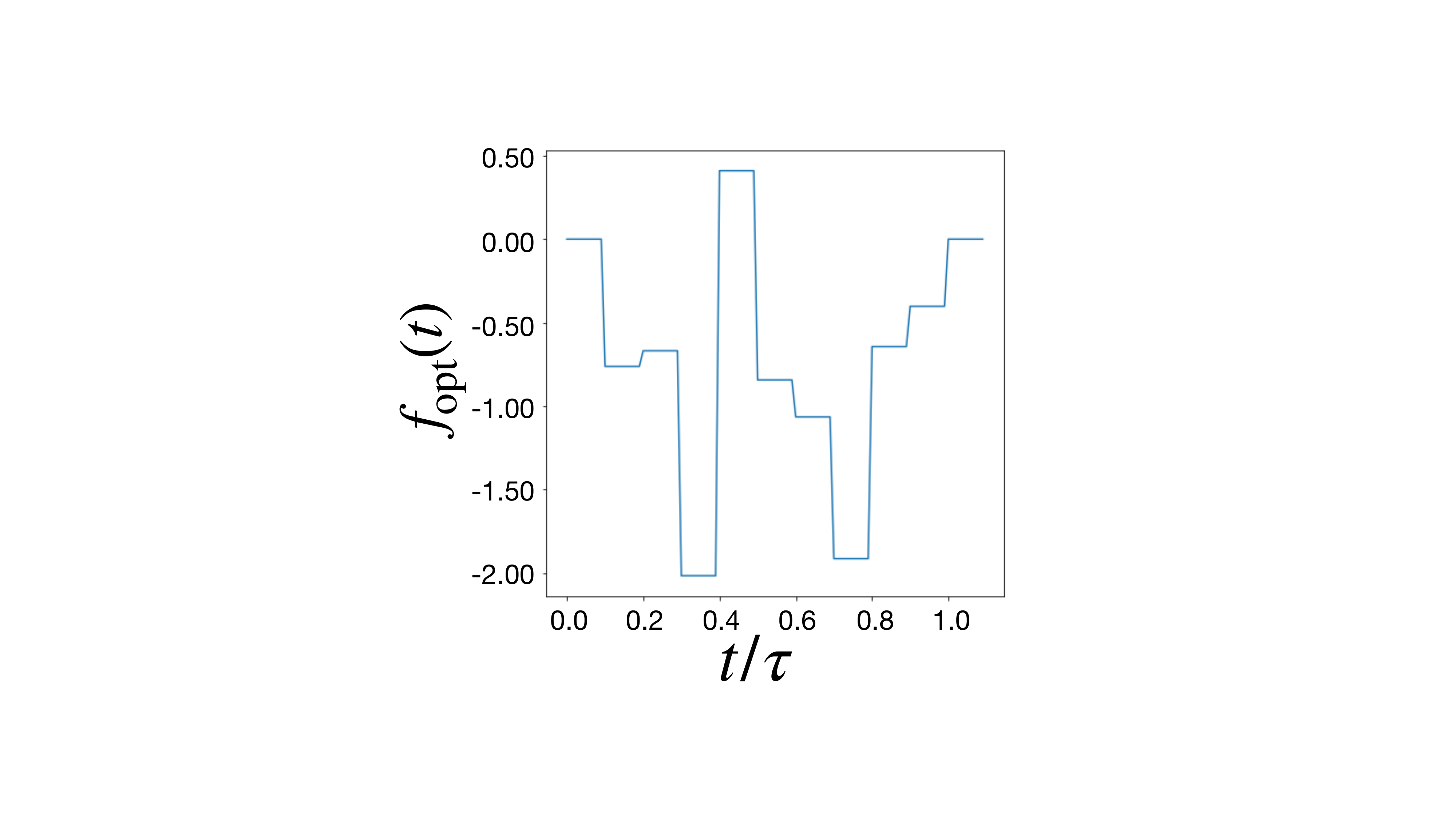}
		\caption{We show the form taken by $f_{opt}(t)$ for two different runs of the optimization process. Although they both reduce the entropy production of approximately the same amount (99.86\% in panel {\bf (a)} and 99.80\% in panel {\bf (b)} respectively), the trends followed by the control function are visibly different.}\label{Figura:ex1}
	\end{figure}
	
	We start with the first approach, introduced in the previous section, that aims to reduce the relative entropy. 
	We introduce the entropy production reduction
$		\Delta \Sigma = 1-{\Sigma_{opt}}/{\Sigma_{free}}$,
	where $\Sigma_{opt}$ is given by Eq.~\eqref{rel_entr} and $\Sigma_{free}$ is associated to the case without optimization term in the Hamiltonian. Likewise, the reduction of the work done on the systems is
$	\Delta W = 1-({\Delta U_{opt}+E_{in}})/{\Delta U_{free}}$,
	where $E_{in}$ is an estimation of the energy spent for the optimization, defined as \cite{art:rif.15}
	\begin{equation}
	E_{in} = \left\vert\int_{0}^{1}\tr(\rho(t) f_{opt}(t)\sigma_y)dt\right\vert.
	\end{equation}
	and $\Delta U$ is the change of the internal energy $U(t)=\tr(\rho(t) H(t))$ of the system between initial and final state. 
	
		\begin{table}[b!]
		\begin{tabular}{cccccc}
			\hline
			\hline
			$J(t)$& $\Sigma_{free}/\beta$ & $\Sigma_{opt}/\beta$ & $\Delta U_\text{free}$ & $\Delta U_\text{opt}$ &  $E_{in}$\\
			\hline
			\hline
			$J_1(t)$& $0.600644$ & $0.367022$ & $0$ & $-0.233621$ &   $0.005392$ \\
			$J_2(t)$& $0.575289$ & $0.366835$ & $-0.025354$ & $-0.233808$ &  $0.004581$ \\
			\hline
		\end{tabular}
		\caption{Results for a simulation of the free and optimized evolution for the different choices of $J(t)$ in the text. The optimized quantities are very close for both the different interaction term functions. In both cases, we achieved an error of $10^{-6}$.}\label{tab:1}
		\end{table}
	
	As our control process starts only after a measurement, the second approach to the quantification of irreversibility gives additional information about the system. Our $f_{opt}(t)$ then depends on the initial state. Based on our knowledge of the initial pure state of the system, we want the final state as close as possible to the adiabatic one (that is, the corresponding eigenvector of $H_S(\tau)$). Therefore, our performance measure for this approach will be the fidelity of the final state with the adiabatic target $|\langle \phi(\tau)|\phi_{ad}(\tau) \rangle|^2$. For the third approach, we solve the previous problem, this time with a LSTM Neural Network, as discussed in {\bf Physical system and methodology}.
	
	\begin{figure}[t!]
{\bf (a)}\hskip3.8cm{\bf (b)}\\
		\includegraphics[width=0.5\columnwidth]{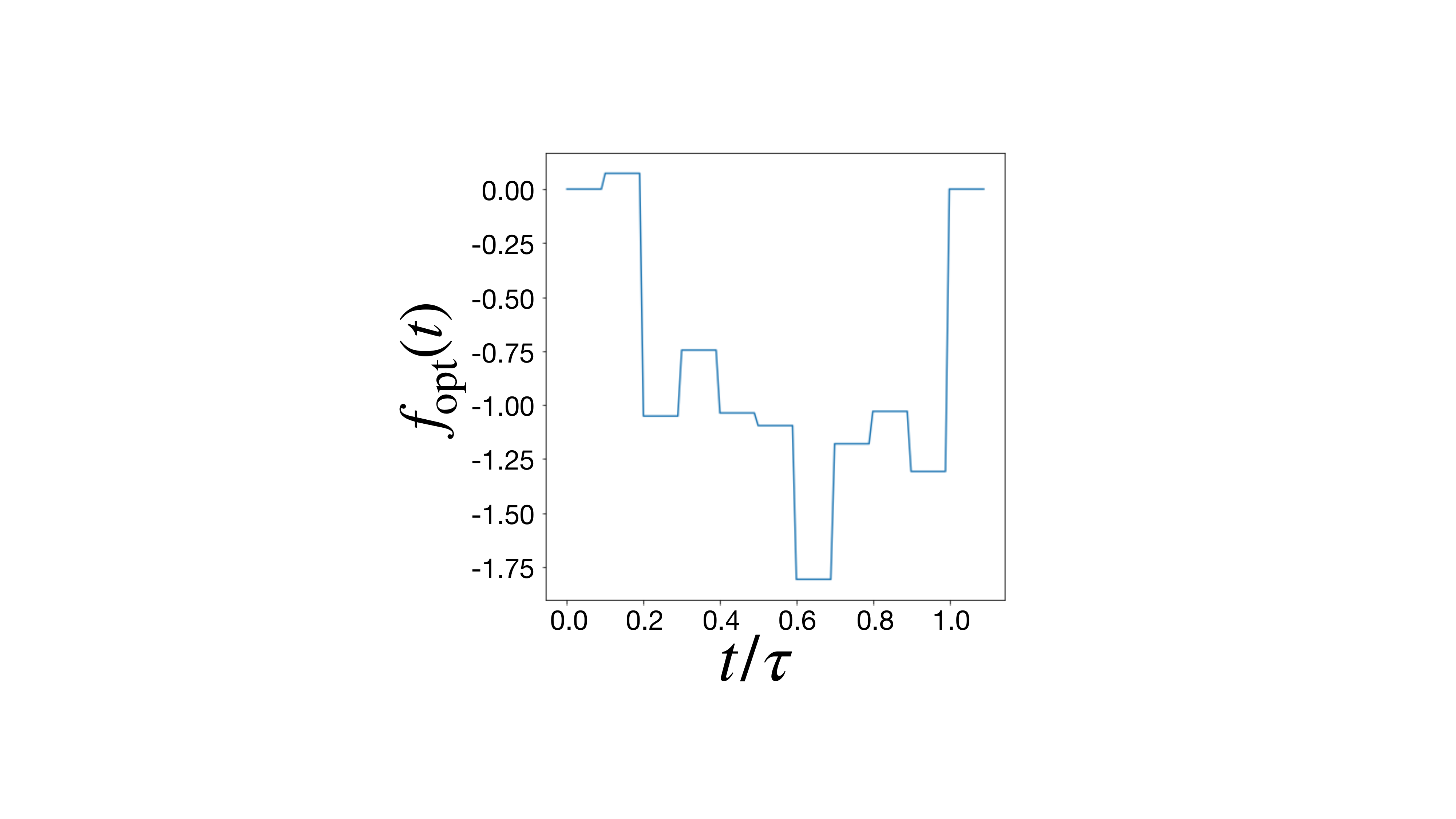}~\includegraphics[width=0.5\columnwidth]{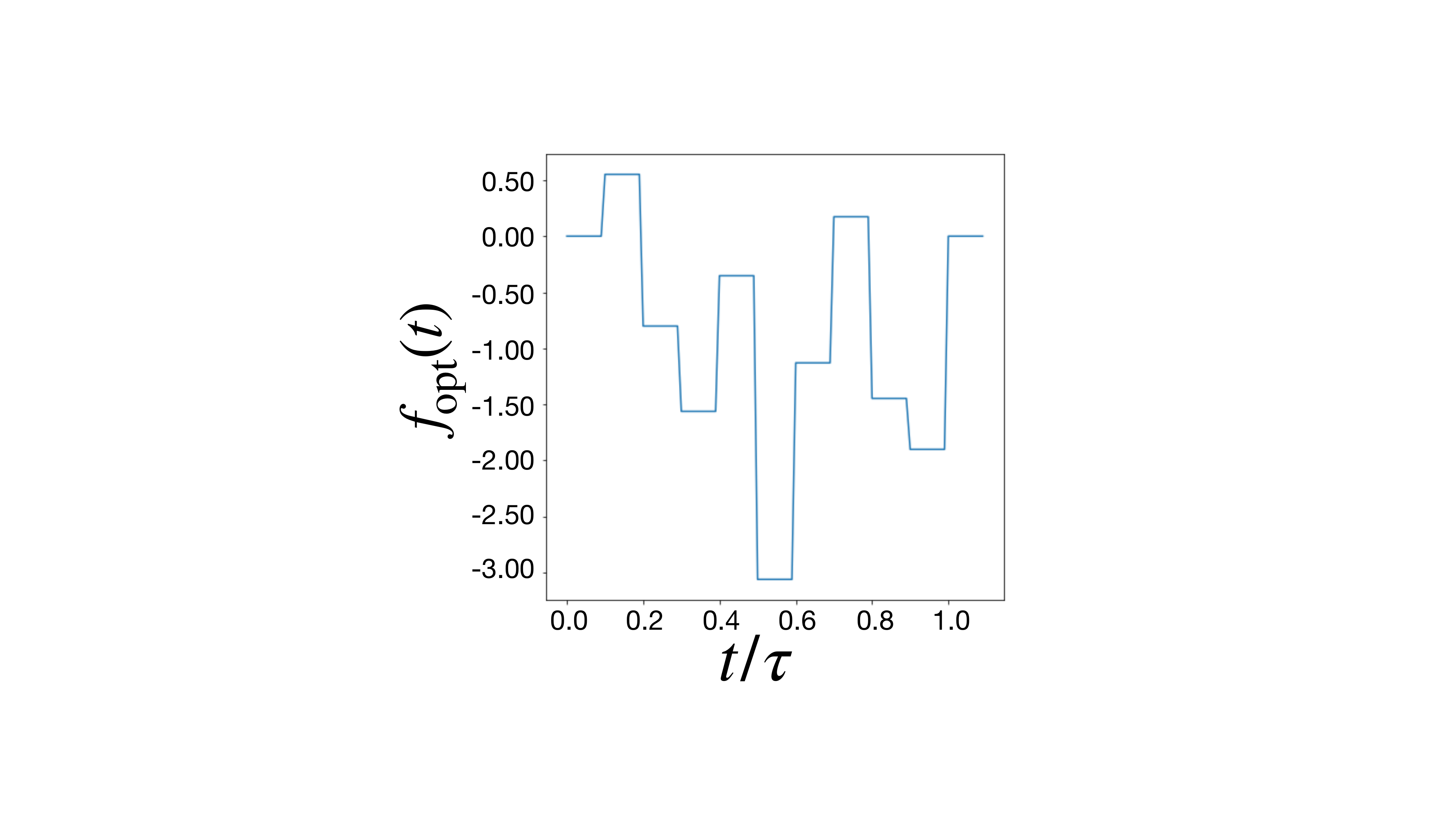}
		\caption{Panel {\bf (a)} [{\bf (b)}]: example of $f_{opt}(t)$ for an initial $\ket{\uparrow}$ [$\ket{\downarrow}$] state of $H_S(0)$. Here, $\sigma_z\ket{\uparrow}=\ket{\uparrow}$ while $\sigma_z\ket{\downarrow}=-\ket{\downarrow}$. The corresponding fidelity with the targets is $0.997$.}\label{Figura:ex2}
	\end{figure}
	
	We divided the dynamics of our system in $10$ steps and set $\mu^*=3$. We considered $B_x(t)=B_0\sin(\frac{\pi t}{2\tau})$ in Eq.~\eqref{1spin}. For each of the RL methods considered here, we ran 20 simulations of a training consisting in 300 epochs with a batch of 30 systems. In Fig.~\ref{Figura:learning} we show a typical example of a learning curve with successful training. Using the first approach with an initial thermal state with $\beta=1$, we successfully reduced both the relative entropy $\Delta \Sigma=(99\pm1)\%$ and the work done on the system $\Delta W=(91\pm9)\%$. Examples of $f_{opt}(t)$ are given in Fig.~\ref{Figura:ex1}.
	When the second approach to irreversibility was used, we obtained the fidelity with the adiabatic target $F_{ad}(\tau)=|\langle \phi(\tau)|\phi_{ad}(\tau) \rangle|^2=0.997\pm0.002$. In Fig.~\ref{Figura:ex2} we show an example of $f_{opt}(t)$ for this case.
	Finally, for the third approach we obtained $F_{ad}(\tau)=0.998\pm0.001$.
	
\begin{figure}[b!]
		{\bf (a)}\hskip3.8cm{\bf (b)}\\
\includegraphics[width=0.5\columnwidth]{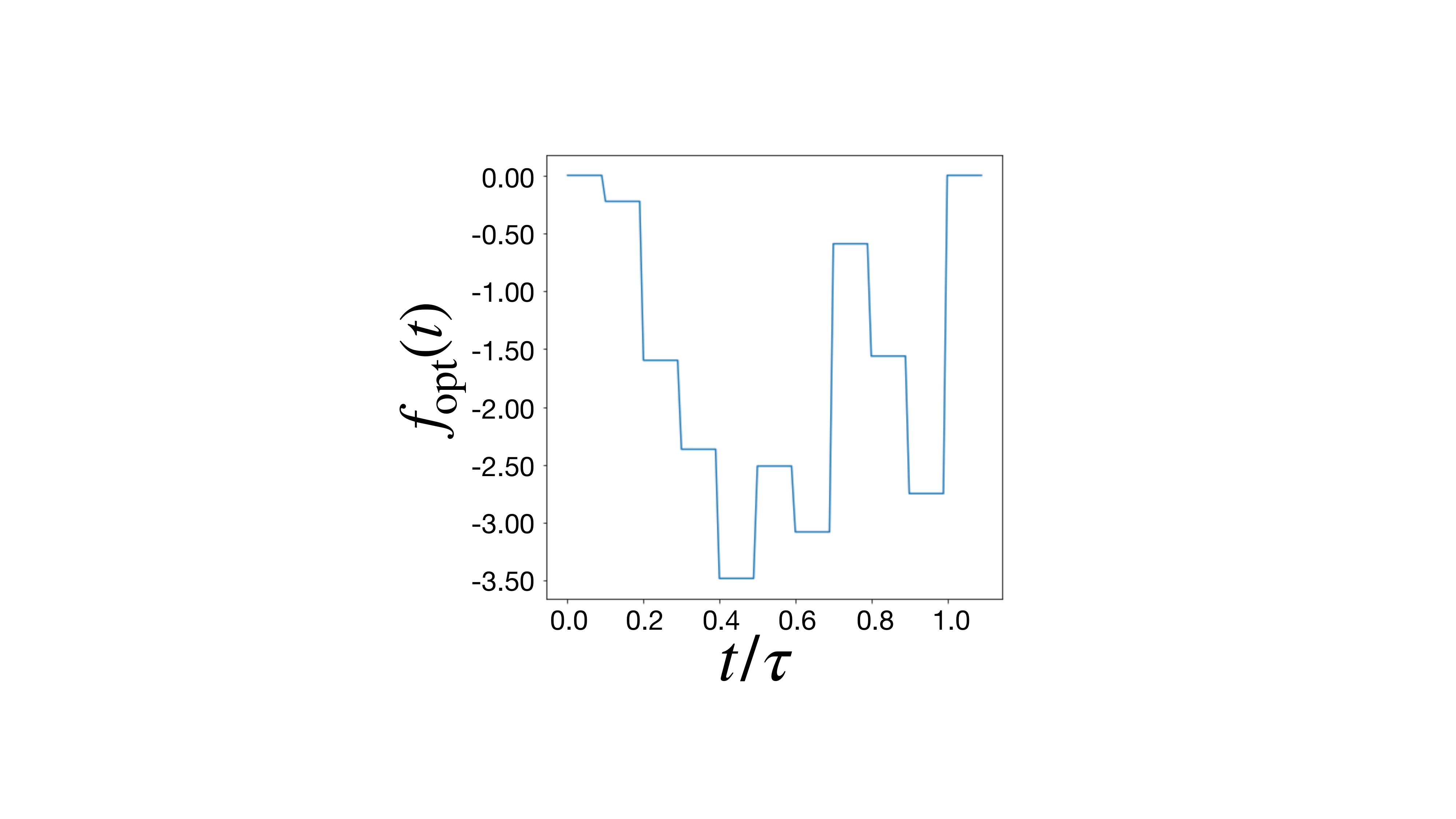}~\includegraphics[width=0.5\columnwidth]{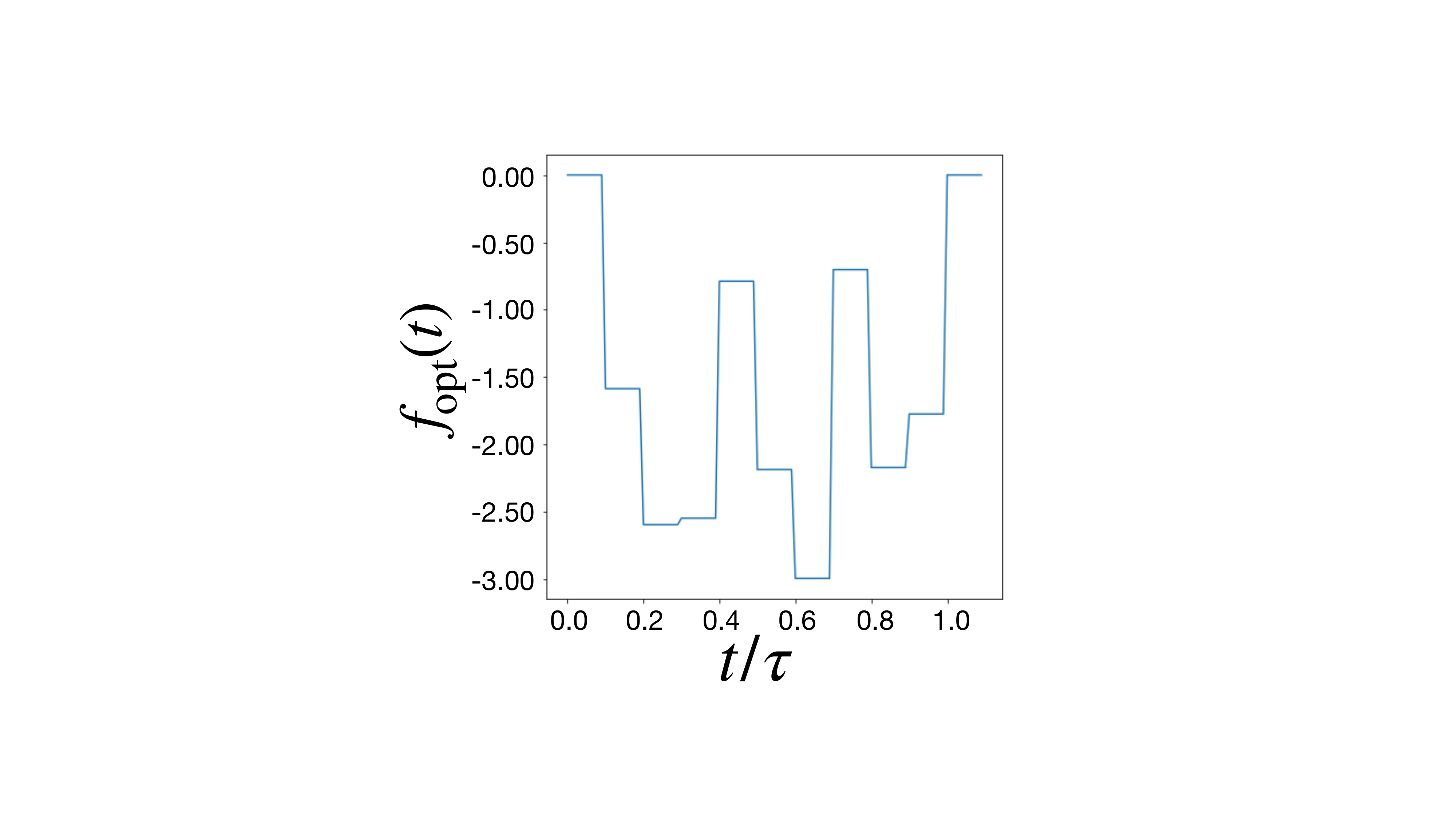}
		\caption{Panel {\bf (a)} [Panel {\bf (b)}] example of $f_{opt}(t)$ for an initial $\ket{\uparrow}$ [$\ket{\downarrow}$] state of $H_S(0)$. Here $B_x(t)=B_0\sin[({\pi}/{2})\sin({\pi t}/{2\tau})^2]$.}\label{Figura:exb}
	\end{figure}

	We have rounded our analysis by running a single simulation for a different choice of time-dependent field, namely $B_x(t)=B_0\sin[\frac{\pi}{2}\sin^2(\frac{\pi t}{2\tau})]$, obtaining a value of the adiabatic fidelity as large as $F_{ad}(\tau)\approx0.998$. The corresponding functions $f_{opt}(t)$ are shown in Fig.~\ref{Figura:exb}.
	
\noindent
{\it Time-dependent coupling of spin-$1/2$ particles}: 
	We now consider a slightly more complicated system composed of two two-level systems with Hamiltonian
	\begin{equation}
	H_S(t)=\sigma_z^1+\frac{1}{2}\sigma_z^2+J(t)({\sigma_{x}^1}{\sigma_{x}^2}+{\sigma_{y}^1}{\sigma_{y}^2}),
	\end{equation}
	where the coupling strength $J(t)$ evolves in the time interval $t \in [0,\tau]$. We start with both spins in a termal state with an inverse temperature $\beta$. Our control term is
	\begin{equation}
	H_{opt}(t)={f_{opt}(t)}({\sigma_{x}^1}{\sigma_{y}^2}-{\sigma_{y}^2}{\sigma_{x}^1})/2.
	\end{equation}
	\begin{figure}[t!]
	{\bf (a)}\hskip3.8cm{\bf (b)}\\
		\includegraphics[width=0.5\columnwidth]{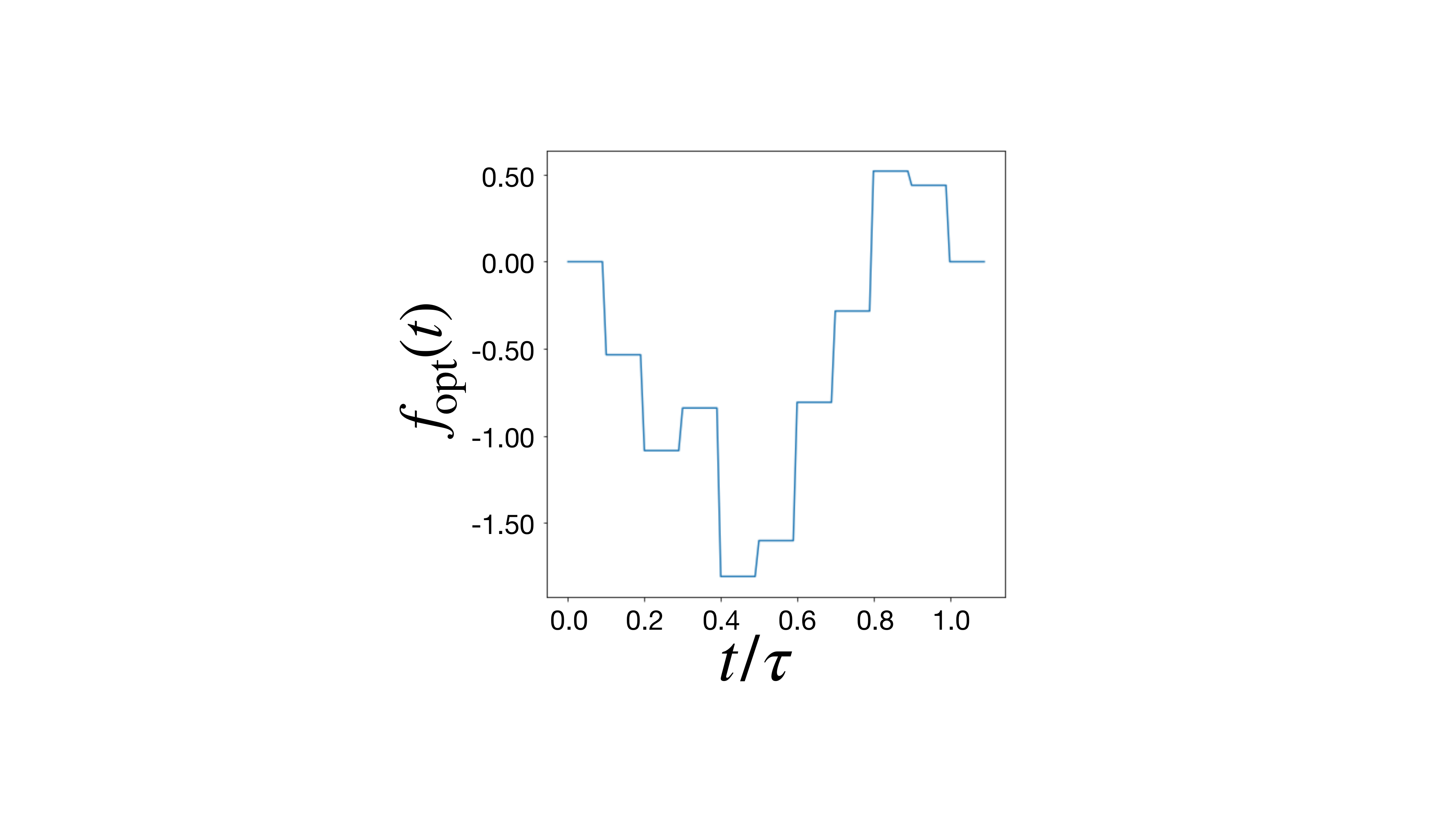}~\includegraphics[width=0.5\columnwidth]{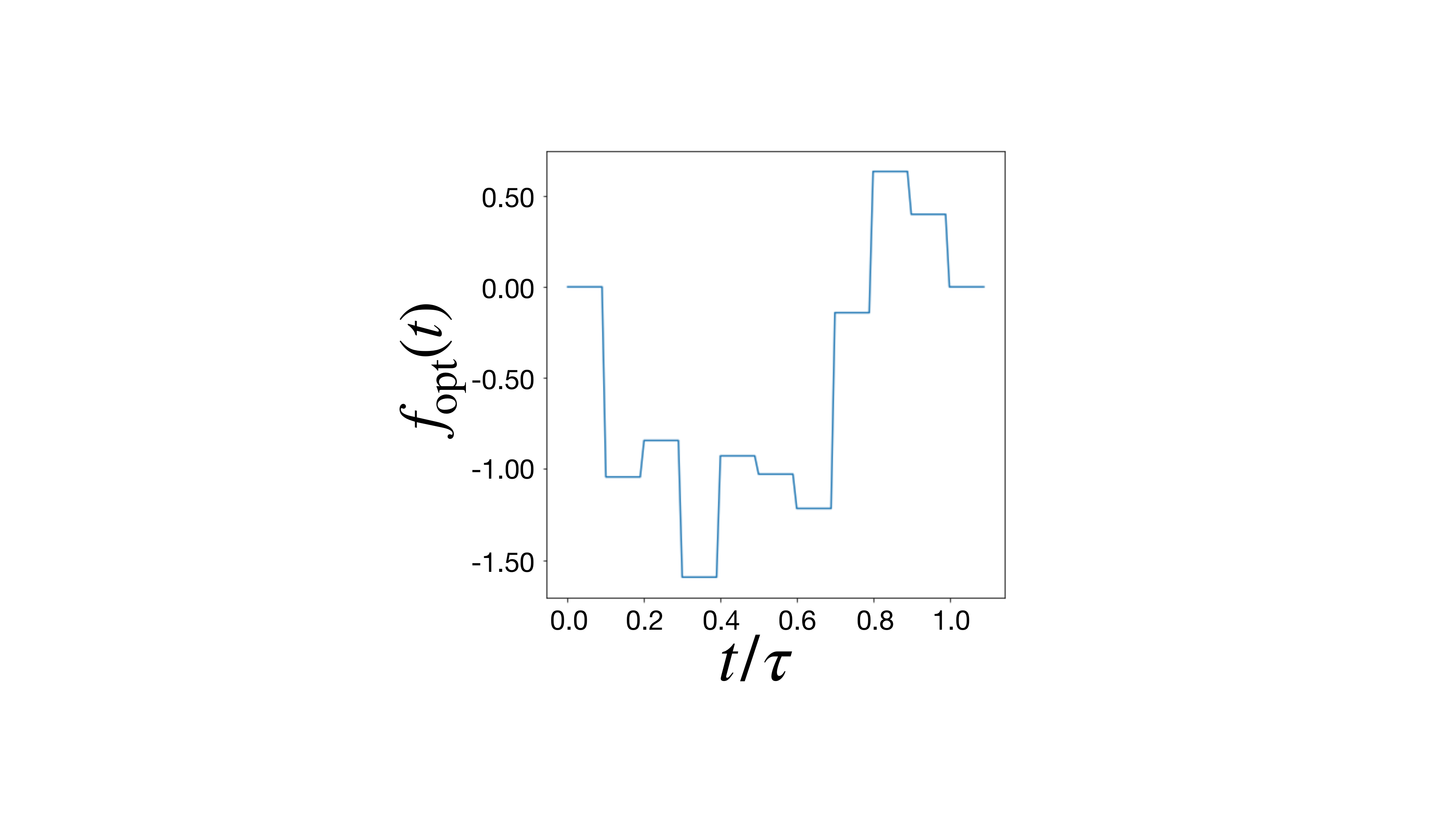}
		\caption{We plot $f_{opt}(t)$ for the time-dependent coupling rates $J_1(t)$
		[panel {\bf (a)}] and $J_2(t)$ 
		[panel {\bf (b)}].}\label{Figura:f2sp}
	\end{figure}
	We aim at minimizing Eq.~\eqref{rel_entr}, this time using a LSTM Neural Network, as described in {\bf Physical system and methodology}. As the variation in the free energy between the initial and final state~\cite{art:rif.2, art:rif.5, art:rif.6}
$\Delta F= \Delta U-{\Sigma}/{\beta}$
	for both the free and the optimized process must be the same, we set the error in our energy measurements to be the difference in this quantity for the two processes.
	
	We ran a simulation where the dynamics of our system is divided in $10$ steps and took $\mu^*=3$. We used a batch of $30$ systems and considered $100$ epochs, choosing the time-dependent coupling rate 
$	J_1(t)=\chi({t}/{\tau}-0.5)$
with $\chi(t-t_0)=1$ at $t=t_0$ and being null otherwise. We have also considered 
$J_2(t)=\sin[{\pi}/{2}-({\pi}/{2})\cos(\frac{\pi t}{2\tau})]$,
	both for an initial thermal state with $\beta=1$. Our results are shown in Fig.~\ref{Figura:f2sp} and Table~\ref{tab:1}. A successful reduction of entropy production  is achieved in both cases. Moreover, the entropy production $\Sigma_{opt}$ for both optimized processes takes very similar values. Similar considerations hold for $\Delta U_{opt}$. This is encouraging, although not surprising, as for both processes we have $J(0)=0$ and $J(\tau)=1$ so that the corresponding adiabatic process is the same, and we have in fact the same $\Delta F$. 
	
	Next, using $J_1(t)$, we changed the temperature of the initial state of the system in the range $\beta\in[0.1,2.1]$, dividing this interval in $20$ steps. Running a single simulation for each value of $\beta$, we obtained a mean entropy production reduction $\Delta\Sigma\approx 36\%$ in this interval.
	
\noindent
{\bf Conclusions.--}
We have proposed and benchmarked a technique a deep RL-based approach to reduce the degree of irreversibility resulting from a non-equilibrium thermodynamic transformation of a closed quantum system. 
Our method can be used with an arbitrary choice of the function characterizing the dissipative process undergone by the system and requires little knowledge of the system dynamics. Moreover, it can be applied without tracking the state of the system during the evolution, thus potentially easing any experimental implementations. 

We applied our technique to two simple yet relevant problems: we successfully reduced the entropy production and the distance of the final state from the adiabatic target for a spin-$1/2$ particle subjected to a time-dependent magnetic field and the entropy production resulting from the time-dependent coupling between two spin-$1/2$ particles. While we focused on simple models, it would be interesting to apply the proposed approach to many-body quantum systems. This could help significantly in the development of an efficient mesoscopic thermal engine operating under realistic conditions.
In particular, our third approach 
could be advantageous when tackling high-dimensional systems. However, such systems will still embody a challenge, as they could require a large number of control terms, and the optimization would still suffer from the increase of dimensionality of the agent actions space. This could lead to a decrease in the performances or at least to an increase in the number of experiments required for a successful optimization, which are issues that we will address in future. A natural further development of our work would be the extension to open quantum systems dynamics.

\acknowledgements
SS thanks the Centre for Theoretical Atomic, Molecular, and Optical Physics, School of Mathematics and Physics, Queen's University Belfast for hospitality during the initial phases of this work. We acknowledge support from the H2020-FETOPEN-2018-2020 TEQ (grant nr. 766900), the DfE-SFI Investigator Programme (grant 15/IA/2864), COST Action CA15220, the Royal Society Wolfson Research Fellowship (RSWF\textbackslash R3\textbackslash183013), the Leverhulme Trust Research Project Grant (grant nr.~RGP-2018-266), the UK EPSRC (grant nr.~EP/T028106/1), and the PRIN project 2017SRN-BRK QUSHIP funded by MIUR.
\nocite{chollet2015keras}
\nocite{adam}
\nocite{Zhang_2019}
\nocite{2020SciPy-NMeth}

\begin{thebibliography}{59}%
\makeatletter
\providecommand \@ifxundefined [1]{%
 \@ifx{#1\undefined}
}%
\providecommand \@ifnum [1]{%
 \ifnum #1\expandafter \@firstoftwo
 \else \expandafter \@secondoftwo
 \fi
}%
\providecommand \@ifx [1]{%
 \ifx #1\expandafter \@firstoftwo
 \else \expandafter \@secondoftwo
 \fi
}%
\providecommand \natexlab [1]{#1}%
\providecommand \enquote  [1]{``#1''}%
\providecommand \bibnamefont  [1]{#1}%
\providecommand \bibfnamefont [1]{#1}%
\providecommand \citenamefont [1]{#1}%
\providecommand \href@noop [0]{\@secondoftwo}%
\providecommand \href [0]{\begingroup \@sanitize@url \@href}%
\providecommand \@href[1]{\@@startlink{#1}\@@href}%
\providecommand \@@href[1]{\endgroup#1\@@endlink}%
\providecommand \@sanitize@url [0]{\catcode `\\12\catcode `\$12\catcode
  `\&12\catcode `\#12\catcode `\^12\catcode `\_12\catcode `\%12\relax}%
\providecommand \@@startlink[1]{}%
\providecommand \@@endlink[0]{}%
\providecommand \url  [0]{\begingroup\@sanitize@url \@url }%
\providecommand \@url [1]{\endgroup\@href {#1}{\urlprefix }}%
\providecommand \urlprefix  [0]{URL }%
\providecommand \Eprint [0]{\href }%
\providecommand \doibase [0]{http://dx.doi.org/}%
\providecommand \selectlanguage [0]{\@gobble}%
\providecommand \bibinfo  [0]{\@secondoftwo}%
\providecommand \bibfield  [0]{\@secondoftwo}%
\providecommand \translation [1]{[#1]}%
\providecommand \BibitemOpen [0]{}%
\providecommand \bibitemStop [0]{}%
\providecommand \bibitemNoStop [0]{.\EOS\space}%
\providecommand \EOS [0]{\spacefactor3000\relax}%
\providecommand \BibitemShut  [1]{\csname bibitem#1\endcsname}%
\let\auto@bib@innerbib\@empty
\bibitem [{\citenamefont {Vinjanampathy}\ and\ \citenamefont
  {Anders}(2016)}]{art:rif.1}%
  \BibitemOpen
  \bibfield  {author} {\bibinfo {author} {\bibfnamefont {S.}~\bibnamefont
  {Vinjanampathy}}\ and\ \bibinfo {author} {\bibfnamefont {J.}~\bibnamefont
  {Anders}},\ }\href {https://doi.org/10.1080/00107514.2016.1201896} {\bibfield
   {journal} {\bibinfo  {journal} {Contemp. Phys. {\bf 57}, 545}\ } (\bibinfo
  {year} {2016})}\BibitemShut {NoStop}%
\bibitem [{\citenamefont {Deffner}\ and\ \citenamefont
  {Campbell}(2019{\natexlab{a}})}]{DeffnerCampbell}%
  \BibitemOpen
  \bibfield  {author} {\bibinfo {author} {\bibfnamefont {S.}~\bibnamefont
  {Deffner}}\ and\ \bibinfo {author} {\bibfnamefont {S.}~\bibnamefont
  {Campbell}},\ }\href {http://dx.doi.org/10.1088/2053-2571/ab21c6} {\emph
  {\bibinfo {title} {Quantum Thermodynamics}}}\ (\bibinfo  {publisher} {Morgan
  and Claypool Publishers},\ \bibinfo {year} {2019})\BibitemShut {NoStop}%
\bibitem [{\citenamefont {Mitchison}(2019)}]{Marksman}%
  \BibitemOpen
  \bibfield  {author} {\bibinfo {author} {\bibfnamefont {M.~T.}\ \bibnamefont
  {Mitchison}},\ }\href {https://doi.org/10.1080/00107514.2019.1631555}
  {\bibfield  {journal} {\bibinfo  {journal} {Contemp. Phys.}\ }\textbf
  {\bibinfo {volume} {60}},\ \bibinfo {pages} {164} (\bibinfo {year}
  {2019})}\BibitemShut {NoStop}%
\bibitem [{\citenamefont {Kosloff}\ and\ \citenamefont {Levy}(2014)}]{Kosloff}%
  \BibitemOpen
  \bibfield  {author} {\bibinfo {author} {\bibfnamefont {R.}~\bibnamefont
  {Kosloff}}\ and\ \bibinfo {author} {\bibfnamefont {A.}~\bibnamefont {Levy}},\
  }\href {https://doi.org/10.1146/annurev-physchem-040513-103724} {\bibfield
  {journal} {\bibinfo  {journal} {Annual Review of Physical Chemistry}\
  }\textbf {\bibinfo {volume} {65}},\ \bibinfo {pages} {365} (\bibinfo {year}
  {2014})}\BibitemShut {NoStop}%
\bibitem [{\citenamefont {Barontini}\ and\ \citenamefont
  {Paternostro}(2019)}]{Barontini2019}%
  \BibitemOpen
  \bibfield  {author} {\bibinfo {author} {\bibfnamefont {G.}~\bibnamefont
  {Barontini}}\ and\ \bibinfo {author} {\bibfnamefont {M.}~\bibnamefont
  {Paternostro}},\ }\href
  {https://iopscience.iop.org/article/10.1088/1367-2630/ab2684} {\bibfield
  {journal} {\bibinfo  {journal} {New Journal of Physics}\ }\textbf {\bibinfo
  {volume} {21}},\ \bibinfo {pages} {063019} (\bibinfo {year}
  {2019})}\BibitemShut {NoStop}%
\bibitem [{\citenamefont {Niedenzu}\ \emph {et~al.}(2018)\citenamefont
  {Niedenzu}, \citenamefont {Mukherjee}, \citenamefont {Ghosh}, \citenamefont
  {Kofman},\ and\ \citenamefont {Kurizki}}]{art:rif.4}%
  \BibitemOpen
  \bibfield  {author} {\bibinfo {author} {\bibfnamefont {W.}~\bibnamefont
  {Niedenzu}}, \bibinfo {author} {\bibfnamefont {V.}~\bibnamefont {Mukherjee}},
  \bibinfo {author} {\bibfnamefont {A.}~\bibnamefont {Ghosh}}, \bibinfo
  {author} {\bibfnamefont {A.~G.}\ \bibnamefont {Kofman}}, \ and\ \bibinfo
  {author} {\bibfnamefont {G.}~\bibnamefont {Kurizki}},\ }\href
  {https://doi.org/10.1038/s41467-017-01991-6} {\bibfield  {journal} {\bibinfo
  {journal} {Nat. Comm. {\bf 9}}\ } (\bibinfo {year} {2018})}\BibitemShut
  {NoStop}%
\bibitem [{\citenamefont {Agarwalla}\ \emph {et~al.}(2017)\citenamefont
  {Agarwalla}, \citenamefont {Jiang},\ and\ \citenamefont
  {Segal}}]{art:rif.10}%
  \BibitemOpen
  \bibfield  {author} {\bibinfo {author} {\bibfnamefont {B.~K.}\ \bibnamefont
  {Agarwalla}}, \bibinfo {author} {\bibfnamefont {J.-H.}\ \bibnamefont
  {Jiang}}, \ and\ \bibinfo {author} {\bibfnamefont {D.}~\bibnamefont
  {Segal}},\ }\href {https://link.aps.org/doi/10.1103/PhysRevB.96.104304}
  {\bibfield  {journal} {\bibinfo  {journal} {Phys. Rev. B {\bf 96}, 104304}\ }
  (\bibinfo {year} {2017})}\BibitemShut {NoStop}%
\bibitem [{\citenamefont {Klaers}\ \emph {et~al.}(2017)\citenamefont {Klaers},
  \citenamefont {Faelt}, \citenamefont {Imamoglu},\ and\ \citenamefont
  {Togan}}]{art:rif.11}%
  \BibitemOpen
  \bibfield  {author} {\bibinfo {author} {\bibfnamefont {J.}~\bibnamefont
  {Klaers}}, \bibinfo {author} {\bibfnamefont {S.}~\bibnamefont {Faelt}},
  \bibinfo {author} {\bibfnamefont {A.}~\bibnamefont {Imamoglu}}, \ and\
  \bibinfo {author} {\bibfnamefont {E.}~\bibnamefont {Togan}},\ }\href
  {https://link.aps.org/doi/10.1103/PhysRevX.7.031044} {\bibfield  {journal}
  {\bibinfo  {journal} {Phys. Rev. X {\bf 7}, 031044}\ } (\bibinfo {year}
  {2017})}\BibitemShut {NoStop}%
\bibitem [{\citenamefont {von Lindenfels}\ \emph {et~al.}(2019)\citenamefont
  {von Lindenfels}, \citenamefont {Gr\"ab}, \citenamefont {Schmiegelow},
  \citenamefont {Kaushal}, \citenamefont {Schulz}, \citenamefont {Mitchison},
  \citenamefont {Goold}, \citenamefont {Schmidt-Kaler},\ and\ \citenamefont
  {Poschinger}}]{Lindenfels2019}%
  \BibitemOpen
  \bibfield  {author} {\bibinfo {author} {\bibfnamefont {D.}~\bibnamefont {von
  Lindenfels}}, \bibinfo {author} {\bibfnamefont {O.}~\bibnamefont {Gr\"ab}},
  \bibinfo {author} {\bibfnamefont {C.~T.}\ \bibnamefont {Schmiegelow}},
  \bibinfo {author} {\bibfnamefont {V.}~\bibnamefont {Kaushal}}, \bibinfo
  {author} {\bibfnamefont {J.}~\bibnamefont {Schulz}}, \bibinfo {author}
  {\bibfnamefont {M.~T.}\ \bibnamefont {Mitchison}}, \bibinfo {author}
  {\bibfnamefont {J.}~\bibnamefont {Goold}}, \bibinfo {author} {\bibfnamefont
  {F.}~\bibnamefont {Schmidt-Kaler}}, \ and\ \bibinfo {author} {\bibfnamefont
  {U.~G.}\ \bibnamefont {Poschinger}},\ }\href
  {https://link.aps.org/doi/10.1103/PhysRevLett.123.080602} {\bibfield
  {journal} {\bibinfo  {journal} {Phys. Rev. Lett.}\ }\textbf {\bibinfo
  {volume} {123}},\ \bibinfo {pages} {080602} (\bibinfo {year}
  {2019})}\BibitemShut {NoStop}%
\bibitem [{\citenamefont {Peterson}\ \emph {et~al.}(2019)\citenamefont
  {Peterson}, \citenamefont {Batalh\~ao}, \citenamefont {Herrera},
  \citenamefont {Souza}, \citenamefont {Sarthour}, \citenamefont {Oliveira},\
  and\ \citenamefont {Serra}}]{Peterson2019}%
  \BibitemOpen
  \bibfield  {author} {\bibinfo {author} {\bibfnamefont {J.~P.~S.}\
  \bibnamefont {Peterson}}, \bibinfo {author} {\bibfnamefont {T.~B.}\
  \bibnamefont {Batalh\~ao}}, \bibinfo {author} {\bibfnamefont
  {M.}~\bibnamefont {Herrera}}, \bibinfo {author} {\bibfnamefont {A.~M.}\
  \bibnamefont {Souza}}, \bibinfo {author} {\bibfnamefont {R.~S.}\ \bibnamefont
  {Sarthour}}, \bibinfo {author} {\bibfnamefont {I.~S.}\ \bibnamefont
  {Oliveira}}, \ and\ \bibinfo {author} {\bibfnamefont {R.~M.}\ \bibnamefont
  {Serra}},\ }\href {https://link.aps.org/doi/10.1103/PhysRevLett.123.240601}
  {\bibfield  {journal} {\bibinfo  {journal} {Phys. Rev. Lett.}\ }\textbf
  {\bibinfo {volume} {123}},\ \bibinfo {pages} {240601} (\bibinfo {year}
  {2019})}\BibitemShut {NoStop}%
\bibitem [{\citenamefont {Landi}\ and\ \citenamefont
  {Paternostro}(2020)}]{entropyproductionRMP}%
  \BibitemOpen
  \bibfield  {author} {\bibinfo {author} {\bibfnamefont {G.~T.}\ \bibnamefont
  {Landi}}\ and\ \bibinfo {author} {\bibfnamefont {M.}~\bibnamefont
  {Paternostro}},\ }\href {https://arxiv.org/abs/2009.07668} {\bibfield
  {journal} {\bibinfo  {journal} {arXiv:2009.07668}\ } (\bibinfo {year}
  {2020})}\BibitemShut {NoStop}%
\bibitem [{\citenamefont {del Campo}\ \emph {et~al.}(2014)\citenamefont {del
  Campo}, \citenamefont {Goold},\ and\ \citenamefont
  {Paternostro}}]{art:rif.13}%
  \BibitemOpen
  \bibfield  {author} {\bibinfo {author} {\bibfnamefont {A.}~\bibnamefont {del
  Campo}}, \bibinfo {author} {\bibfnamefont {J.}~\bibnamefont {Goold}}, \ and\
  \bibinfo {author} {\bibfnamefont {M.}~\bibnamefont {Paternostro}},\ }\href
  {https://doi.org/10.1038/srep06208} {\bibfield  {journal} {\bibinfo
  {journal} {Sci. Rep.}\ }\textbf {\bibinfo {volume} {4}},\ \bibinfo {pages}
  {6208} (\bibinfo {year} {2014})}\BibitemShut {NoStop}%
\bibitem [{\citenamefont {Torrontegui}\ \emph {et~al.}(2013)\citenamefont
  {Torrontegui}, \citenamefont {Ib{\'a}{\~n}ez}, \citenamefont
  {Mart{\'i}nez-Garaot}, \citenamefont {Modugno}, \citenamefont {del Campo},
  \citenamefont {Gu{\'e}ry-Odelin}, \citenamefont {Ruschhaupt}, \citenamefont
  {Chen},\ and\ \citenamefont {Muga}}]{TORRONTEGUI2013117}%
  \BibitemOpen
  \bibfield  {author} {\bibinfo {author} {\bibfnamefont {E.}~\bibnamefont
  {Torrontegui}}, \bibinfo {author} {\bibfnamefont {S.}~\bibnamefont
  {Ib{\'a}{\~n}ez}}, \bibinfo {author} {\bibfnamefont {S.}~\bibnamefont
  {Mart{\'i}nez-Garaot}}, \bibinfo {author} {\bibfnamefont {M.}~\bibnamefont
  {Modugno}}, \bibinfo {author} {\bibfnamefont {A.}~\bibnamefont {del Campo}},
  \bibinfo {author} {\bibfnamefont {D.}~\bibnamefont {Gu{\'e}ry-Odelin}},
  \bibinfo {author} {\bibfnamefont {A.}~\bibnamefont {Ruschhaupt}}, \bibinfo
  {author} {\bibfnamefont {X.}~\bibnamefont {Chen}}, \ and\ \bibinfo {author}
  {\bibfnamefont {J.~G.}\ \bibnamefont {Muga}},\ }\href
  {http://www.sciencedirect.com/science/article/pii/B9780124080904000025}
  {\emph {\bibinfo {title} {Adv. At. Mol. Opt. Phys.}}},\ edited by\ \bibinfo
  {editor} {\bibfnamefont {E.}~\bibnamefont {Arimondo}}, \bibinfo {editor}
  {\bibfnamefont {P.~R.}\ \bibnamefont {Berman}}, \ and\ \bibinfo {editor}
  {\bibfnamefont {C.~C.}\ \bibnamefont {Lin}},\ Vol.~\bibinfo {volume} {62}\
  (\bibinfo  {publisher} {Academic Press},\ \bibinfo {year} {2013})\ p.\
  \bibinfo {pages} {117}\BibitemShut {NoStop}%
\bibitem [{\citenamefont {Berry}(2009)}]{art:rif.16}%
  \BibitemOpen
  \bibfield  {author} {\bibinfo {author} {\bibfnamefont {M.~V.}\ \bibnamefont
  {Berry}},\ }\href {https://doi.org/10.1088/1751-8113/42/36/365303} {\bibfield
   {journal} {\bibinfo  {journal} {J. Phys. A: Math. Theor.}\ }\textbf
  {\bibinfo {volume} {\bf 42}},\ \bibinfo {pages} {365303} (\bibinfo {year}
  {2009})}\BibitemShut {NoStop}%
\bibitem [{\citenamefont {Gu\'ery-Odelin}\ \emph {et~al.}(2019)\citenamefont
  {Gu\'ery-Odelin}, \citenamefont {Ruschhaupt}, \citenamefont {Kiely},
  \citenamefont {Torrontegui}, \citenamefont {Mart\'{\i}nez-Garaot},\ and\
  \citenamefont {Muga}}]{RevModPhys.91.045001}%
  \BibitemOpen
  \bibfield  {author} {\bibinfo {author} {\bibfnamefont {D.}~\bibnamefont
  {Gu\'ery-Odelin}}, \bibinfo {author} {\bibfnamefont {A.}~\bibnamefont
  {Ruschhaupt}}, \bibinfo {author} {\bibfnamefont {A.}~\bibnamefont {Kiely}},
  \bibinfo {author} {\bibfnamefont {E.}~\bibnamefont {Torrontegui}}, \bibinfo
  {author} {\bibfnamefont {S.}~\bibnamefont {Mart\'{\i}nez-Garaot}}, \ and\
  \bibinfo {author} {\bibfnamefont {J.~G.}\ \bibnamefont {Muga}},\ }\href
  {\doibase 10.1103/RevModPhys.91.045001} {\bibfield  {journal} {\bibinfo
  {journal} {Rev. Mod. Phys.}\ }\textbf {\bibinfo {volume} {91}},\ \bibinfo
  {pages} {045001} (\bibinfo {year} {2019})}\BibitemShut {NoStop}%
\bibitem [{\citenamefont {Palmero}\ \emph {et~al.}(2016)\citenamefont
  {Palmero}, \citenamefont {Wang}, \citenamefont {Gu{\'{e}}ry-Odelin},
  \citenamefont {Li},\ and\ \citenamefont {Muga}}]{Palmero_2016}%
  \BibitemOpen
  \bibfield  {author} {\bibinfo {author} {\bibfnamefont {M.}~\bibnamefont
  {Palmero}}, \bibinfo {author} {\bibfnamefont {S.}~\bibnamefont {Wang}},
  \bibinfo {author} {\bibfnamefont {D.}~\bibnamefont {Gu{\'{e}}ry-Odelin}},
  \bibinfo {author} {\bibfnamefont {J.-S.}\ \bibnamefont {Li}}, \ and\ \bibinfo
  {author} {\bibfnamefont {J.~G.}\ \bibnamefont {Muga}},\ }\href {\doibase
  10.1088/1367-2630/18/4/043014} {\bibfield  {journal} {\bibinfo  {journal}
  {New Journal of Physics}\ }\textbf {\bibinfo {volume} {18}},\ \bibinfo
  {pages} {043014} (\bibinfo {year} {2016})}\BibitemShut {NoStop}%
\bibitem [{\citenamefont {Mortensen}\ \emph {et~al.}(2018)\citenamefont
  {Mortensen}, \citenamefont {S{\o}rensen}, \citenamefont {M{\o}lmer},\ and\
  \citenamefont {Sherson}}]{Mortensen_2018}%
  \BibitemOpen
  \bibfield  {author} {\bibinfo {author} {\bibfnamefont {H.~L.}\ \bibnamefont
  {Mortensen}}, \bibinfo {author} {\bibfnamefont {J.~J. W.~H.}\ \bibnamefont
  {S{\o}rensen}}, \bibinfo {author} {\bibfnamefont {K.}~\bibnamefont
  {M{\o}lmer}}, \ and\ \bibinfo {author} {\bibfnamefont {J.~F.}\ \bibnamefont
  {Sherson}},\ }\href {\doibase 10.1088/1367-2630/aaac8a} {\bibfield  {journal}
  {\bibinfo  {journal} {New J. Phys.}\ }\textbf {\bibinfo {volume} {20}},\
  \bibinfo {pages} {025009} (\bibinfo {year} {2018})}\BibitemShut {NoStop}%
\bibitem [{\citenamefont {Deng}\ \emph
  {et~al.}(2018{\natexlab{a}})\citenamefont {Deng}, \citenamefont {Diao},
  \citenamefont {Yu}, \citenamefont {del Campo},\ and\ \citenamefont
  {Wu}}]{PhysRevA.97.013628}%
  \BibitemOpen
  \bibfield  {author} {\bibinfo {author} {\bibfnamefont {S.}~\bibnamefont
  {Deng}}, \bibinfo {author} {\bibfnamefont {P.}~\bibnamefont {Diao}}, \bibinfo
  {author} {\bibfnamefont {Q.}~\bibnamefont {Yu}}, \bibinfo {author}
  {\bibfnamefont {A.}~\bibnamefont {del Campo}}, \ and\ \bibinfo {author}
  {\bibfnamefont {H.}~\bibnamefont {Wu}},\ }\href {\doibase
  10.1103/PhysRevA.97.013628} {\bibfield  {journal} {\bibinfo  {journal} {Phys.
  Rev. A}\ }\textbf {\bibinfo {volume} {97}},\ \bibinfo {pages} {013628}
  (\bibinfo {year} {2018}{\natexlab{a}})}\BibitemShut {NoStop}%
\bibitem [{\citenamefont {Ren}\ \emph {et~al.}(2017)\citenamefont {Ren},
  \citenamefont {Wang},\ and\ \citenamefont {Gu}}]{REN201770}%
  \BibitemOpen
  \bibfield  {author} {\bibinfo {author} {\bibfnamefont {F.-H.}\ \bibnamefont
  {Ren}}, \bibinfo {author} {\bibfnamefont {Z.-M.}\ \bibnamefont {Wang}}, \
  and\ \bibinfo {author} {\bibfnamefont {Y.-J.}\ \bibnamefont {Gu}},\ }\href
  {\doibase https://doi.org/10.1016/j.physleta.2016.10.041} {\bibfield
  {journal} {\bibinfo  {journal} {Physics Letters A}\ }\textbf {\bibinfo
  {volume} {381}},\ \bibinfo {pages} {70 } (\bibinfo {year}
  {2017})}\BibitemShut {NoStop}%
\bibitem [{\citenamefont {Saberi}\ \emph {et~al.}(2014)\citenamefont {Saberi},
  \citenamefont {Opatrny}, \citenamefont {M{\o}lmer},\ and\ \citenamefont {del
  Campo}}]{articlead}%
  \BibitemOpen
  \bibfield  {author} {\bibinfo {author} {\bibfnamefont {H.}~\bibnamefont
  {Saberi}}, \bibinfo {author} {\bibfnamefont {T.}~\bibnamefont {Opatrny}},
  \bibinfo {author} {\bibfnamefont {K.}~\bibnamefont {M{\o}lmer}}, \ and\
  \bibinfo {author} {\bibfnamefont {A.}~\bibnamefont {del Campo}},\ }\href
  {10.1103/PhysRevA.90.060301} {\bibfield  {journal} {\bibinfo  {journal}
  {Physical Review A}\ }\textbf {\bibinfo {volume} {90}},\ \bibinfo {pages} 060301(R) (\bibinfo {year}
  {2014})}\BibitemShut {NoStop}%
\bibitem [{\citenamefont {Campbell}\ \emph {et~al.}(2015)\citenamefont
  {Campbell}, \citenamefont {De~Chiara}, \citenamefont {Paternostro},
  \citenamefont {Palma},\ and\ \citenamefont {Fazio}}]{PhysRevLett.114.177206}%
  \BibitemOpen
  \bibfield  {author} {\bibinfo {author} {\bibfnamefont {S.}~\bibnamefont
  {Campbell}}, \bibinfo {author} {\bibfnamefont {G.}~\bibnamefont {De~Chiara}},
  \bibinfo {author} {\bibfnamefont {M.}~\bibnamefont {Paternostro}}, \bibinfo
  {author} {\bibfnamefont {G.~M.}\ \bibnamefont {Palma}}, \ and\ \bibinfo
  {author} {\bibfnamefont {R.}~\bibnamefont {Fazio}},\ }\href {\doibase
  10.1103/PhysRevLett.114.177206} {\bibfield  {journal} {\bibinfo  {journal}
  {Phys. Rev. Lett.}\ }\textbf {\bibinfo {volume} {114}},\ \bibinfo {pages}
  {177206} (\bibinfo {year} {2015})}\BibitemShut {NoStop}%
\bibitem [{\citenamefont {Abah}\ and\ \citenamefont {Lutz}(2017)}]{art:rif.15}%
  \BibitemOpen
  \bibfield  {author} {\bibinfo {author} {\bibfnamefont {O.}~\bibnamefont
  {Abah}}\ and\ \bibinfo {author} {\bibfnamefont {E.}~\bibnamefont {Lutz}},\
  }\href {https://doi.org/10.1209/0295-5075/118/40005} {\bibfield  {journal}
  {\bibinfo  {journal} {EPL (Europhysics Letters)}\ }\textbf {\bibinfo {volume}
  {118}},\ \bibinfo {pages} {40005} (\bibinfo {year} {2017})}\BibitemShut
  {NoStop}%
\bibitem [{\citenamefont {Deng}\ \emph
  {et~al.}(2018{\natexlab{b}})\citenamefont {Deng}, \citenamefont {Chenu},
  \citenamefont {Diao}, \citenamefont {Li}, \citenamefont {Yu}, \citenamefont
  {Coulamy}, \citenamefont {del Campo},\ and\ \citenamefont
  {Wu}}]{Dengeaar5909}%
  \BibitemOpen
  \bibfield  {author} {\bibinfo {author} {\bibfnamefont {S.}~\bibnamefont
  {Deng}}, \bibinfo {author} {\bibfnamefont {A.}~\bibnamefont {Chenu}},
  \bibinfo {author} {\bibfnamefont {P.}~\bibnamefont {Diao}}, \bibinfo {author}
  {\bibfnamefont {F.}~\bibnamefont {Li}}, \bibinfo {author} {\bibfnamefont
  {S.}~\bibnamefont {Yu}}, \bibinfo {author} {\bibfnamefont {I.}~\bibnamefont
  {Coulamy}}, \bibinfo {author} {\bibfnamefont {A.}~\bibnamefont {del Campo}},
  \ and\ \bibinfo {author} {\bibfnamefont {H.}~\bibnamefont {Wu}},\ }\href
  {https://advances.sciencemag.org/content/4/4/eaar5909} {\bibfield  {journal}
  {\bibinfo  {journal} {Sci. Adv.}\ }\textbf {\bibinfo {volume} {4}} (\bibinfo
  {year} {2018}{\natexlab{b}})}\BibitemShut {NoStop}%
\bibitem [{\citenamefont {Abah}\ and\ \citenamefont
  {Paternostro}(2019)}]{PhysRevE.99.022110}%
  \BibitemOpen
  \bibfield  {author} {\bibinfo {author} {\bibfnamefont {O.}~\bibnamefont
  {Abah}}\ and\ \bibinfo {author} {\bibfnamefont {M.}~\bibnamefont
  {Paternostro}},\ }\href {\doibase 10.1103/PhysRevE.99.022110} {\bibfield
  {journal} {\bibinfo  {journal} {Phys. Rev. E}\ }\textbf {\bibinfo {volume}
  {99}},\ \bibinfo {pages} {022110} (\bibinfo {year} {2019})}\BibitemShut
  {NoStop}%
\bibitem [{\citenamefont {\c{C}akmak}\ and\ \citenamefont {M\"ustecapl\ifmmode
  \imath \else \i \fi{}o\ifmmode~\breve{g}\else
  \u{g}\fi{}lu}(2019)}]{PhysRevE.99.032108}%
  \BibitemOpen
  \bibfield  {author} {\bibinfo {author} {\bibfnamefont {B.}~\bibnamefont
  {\c{C}akmak}}\ and\ \bibinfo {author} {\bibfnamefont {{\"O}.~E.}\
  \bibnamefont {M\"ustecapl\ifmmode \imath \else \i
  \fi{}o\ifmmode~\breve{g}\else \u{g}\fi{}lu}},\ }\href {\doibase
  10.1103/PhysRevE.99.032108} {\bibfield  {journal} {\bibinfo  {journal} {Phys.
  Rev. E}\ }\textbf {\bibinfo {volume} {99}},\ \bibinfo {pages} {032108}
  (\bibinfo {year} {2019})}\BibitemShut {NoStop}%
\bibitem [{\citenamefont {Abah}\ and\ \citenamefont
  {Lutz}(2018)}]{PhysRevE.98.032121}%
  \BibitemOpen
  \bibfield  {author} {\bibinfo {author} {\bibfnamefont {O.}~\bibnamefont
  {Abah}}\ and\ \bibinfo {author} {\bibfnamefont {E.}~\bibnamefont {Lutz}},\
  }\href {\doibase 10.1103/PhysRevE.98.032121} {\bibfield  {journal} {\bibinfo
  {journal} {Phys. Rev. E}\ }\textbf {\bibinfo {volume} {98}},\ \bibinfo
  {pages} {032121} (\bibinfo {year} {2018})}\BibitemShut {NoStop}%
\bibitem [{\citenamefont {Abah}\ \emph {et~al.}(2019)\citenamefont {Abah},
  \citenamefont {Paternostro},\ and\ \citenamefont {Lutz}}]{unknown}%
  \BibitemOpen
  \bibfield  {author} {\bibinfo {author} {\bibfnamefont {O.}~\bibnamefont
  {Abah}}, \bibinfo {author} {\bibfnamefont {M.}~\bibnamefont {Paternostro}}, \
  and\ \bibinfo {author} {\bibfnamefont {E.}~\bibnamefont {Lutz}},\ }\href
  {https://arxiv.org/abs/1911.00373} {\bibfield  {journal} {\bibinfo  {journal}
  {arXiv:1911.00373}\ } (\bibinfo {year} {2019})}\BibitemShut {NoStop}%
\bibitem [{\citenamefont {Zheng}\ \emph {et~al.}(2016)\citenamefont {Zheng},
  \citenamefont {Campbell}, \citenamefont {De~Chiara},\ and\ \citenamefont
  {Poletti}}]{Zheng2016}%
  \BibitemOpen
  \bibfield  {author} {\bibinfo {author} {\bibfnamefont {Y.}~\bibnamefont
  {Zheng}}, \bibinfo {author} {\bibfnamefont {S.}~\bibnamefont {Campbell}},
  \bibinfo {author} {\bibfnamefont {G.}~\bibnamefont {De~Chiara}}, \ and\
  \bibinfo {author} {\bibfnamefont {D.}~\bibnamefont {Poletti}},\ }\href
  {\doibase 10.1103/PhysRevA.94.042132} {\bibfield  {journal} {\bibinfo
  {journal} {Phys. Rev. A}\ }\textbf {\bibinfo {volume} {94}},\ \bibinfo
  {pages} {042132} (\bibinfo {year} {2016})}\BibitemShut {NoStop}%
\bibitem [{\citenamefont {Santos}\ and\ \citenamefont
  {Sarandy}(2015)}]{Santos2015}%
  \BibitemOpen
  \bibfield  {author} {\bibinfo {author} {\bibfnamefont {A.~C.}\ \bibnamefont
  {Santos}}\ and\ \bibinfo {author} {\bibfnamefont {M.~S.}\ \bibnamefont
  {Sarandy}},\ }\href {\doibase https://doi.org/10.1038/srep15775} {\bibfield
  {journal} {\bibinfo  {journal} {Sci. Rep.}\ }\textbf {\bibinfo {volume}
  {5}},\ \bibinfo {pages} {15775} (\bibinfo {year} {2015})}\BibitemShut
  {NoStop}%
\bibitem [{\citenamefont {Broecker}\ \emph {et~al.}(2017)\citenamefont
  {Broecker}, \citenamefont {Assaad},\ and\ \citenamefont {Trebst}}]{mlph0}%
  \BibitemOpen
  \bibfield  {author} {\bibinfo {author} {\bibfnamefont {P.}~\bibnamefont
  {Broecker}}, \bibinfo {author} {\bibfnamefont {F.}~\bibnamefont {Assaad}}, \
  and\ \bibinfo {author} {\bibfnamefont {S.}~\bibnamefont {Trebst}},\ }\href
  {https://arxiv.org/abs/1707.00663} {\bibfield  {journal} {\bibinfo  {journal}
  {arXiv:1707.00663}\ } (\bibinfo {year} {2017})}\BibitemShut {NoStop}%
\bibitem [{\citenamefont {Carrasquilla}\ and\ \citenamefont
  {Melko}(2017)}]{mlph1}%
  \BibitemOpen
  \bibfield  {author} {\bibinfo {author} {\bibfnamefont {J.}~\bibnamefont
  {Carrasquilla}}\ and\ \bibinfo {author} {\bibfnamefont {R.~G.}\ \bibnamefont
  {Melko}},\ }\href {https://doi.org/10.1038/nphys4035} {\bibfield  {journal}
  {\bibinfo  {journal} {Nat. Phys.}\ }\textbf {\bibinfo {volume} {13}},\
  \bibinfo {pages} {1745} (\bibinfo {year} {2017})}\BibitemShut {NoStop}%
\bibitem [{\citenamefont {Yoshioka}\ and\ \citenamefont
  {Hamazaki}(2019)}]{PhysRevB.99.214306}%
  \BibitemOpen
  \bibfield  {author} {\bibinfo {author} {\bibfnamefont {N.}~\bibnamefont
  {Yoshioka}}\ and\ \bibinfo {author} {\bibfnamefont {R.}~\bibnamefont
  {Hamazaki}},\ }\href {\doibase 10.1103/PhysRevB.99.214306} {\bibfield
  {journal} {\bibinfo  {journal} {Phys. Rev. B}\ }\textbf {\bibinfo {volume}
  {99}},\ \bibinfo {pages} {214306} (\bibinfo {year} {2019})}\BibitemShut
  {NoStop}%
\bibitem [{\citenamefont {Melnikov}\ \emph {et~al.}(2018)\citenamefont
  {Melnikov}, \citenamefont {Poulsen~Nautrup}, \citenamefont {Krenn},
  \citenamefont {Dunjko}, \citenamefont {Tiersch}, \citenamefont {Zeilinger},\
  and\ \citenamefont {Briegel}}]{Melnikov1221}%
  \BibitemOpen
  \bibfield  {author} {\bibinfo {author} {\bibfnamefont {A.~A.}\ \bibnamefont
  {Melnikov}}, \bibinfo {author} {\bibfnamefont {H.}~\bibnamefont
  {Poulsen~Nautrup}}, \bibinfo {author} {\bibfnamefont {M.}~\bibnamefont
  {Krenn}}, \bibinfo {author} {\bibfnamefont {V.}~\bibnamefont {Dunjko}},
  \bibinfo {author} {\bibfnamefont {M.}~\bibnamefont {Tiersch}}, \bibinfo
  {author} {\bibfnamefont {A.}~\bibnamefont {Zeilinger}}, \ and\ \bibinfo
  {author} {\bibfnamefont {H.~J.}\ \bibnamefont {Briegel}},\ }\href
  {https://www.pnas.org/content/115/6/1221} {\bibfield  {journal} {\bibinfo
  {journal} {Proc. Natl. Acad. Sci.}\ }\textbf {\bibinfo {volume} {115}},\
  \bibinfo {pages} {1221} (\bibinfo {year} {2018})}\BibitemShut {NoStop}%
\bibitem [{\citenamefont {Porotti}\ \emph {et~al.}(2019)\citenamefont
  {Porotti}, \citenamefont {Tamascelli}, \citenamefont {Restelli},\ and\
  \citenamefont {Prati}}]{Porotti2019}%
  \BibitemOpen
  \bibfield  {author} {\bibinfo {author} {\bibfnamefont {R.}~\bibnamefont
  {Porotti}}, \bibinfo {author} {\bibfnamefont {D.}~\bibnamefont {Tamascelli}},
  \bibinfo {author} {\bibfnamefont {M.}~\bibnamefont {Restelli}}, \ and\
  \bibinfo {author} {\bibfnamefont {E.}~\bibnamefont {Prati}},\ }\href
  {https://doi.org/10.1038/s42005-019-0169-x} {\bibfield  {journal} {\bibinfo
  {journal} {Commun. Phys.}\ }\textbf {\bibinfo {volume} {2}},\ \bibinfo
  {pages} {2399} (\bibinfo {year} {2019})}\BibitemShut {NoStop}%
\bibitem [{\citenamefont {Paparelle}\ \emph {et~al.}(2020)\citenamefont
  {Paparelle}, \citenamefont {Moro},\ and\ \citenamefont
  {Prati}}]{PAPARELLE2020126266}%
  \BibitemOpen
  \bibfield  {author} {\bibinfo {author} {\bibfnamefont {I.}~\bibnamefont
  {Paparelle}}, \bibinfo {author} {\bibfnamefont {L.}~\bibnamefont {Moro}}, \
  and\ \bibinfo {author} {\bibfnamefont {E.}~\bibnamefont {Prati}},\ }\href
  {\doibase https://doi.org/10.1016/j.physleta.2020.126266} {\bibfield
  {journal} {\bibinfo  {journal} {Phys. Lett. A}\ }\textbf {\bibinfo {volume}
  {384}},\ \bibinfo {pages} {126266} (\bibinfo {year} {2020})}\BibitemShut
  {NoStop}%
\bibitem [{\citenamefont {Bukov}\ \emph {et~al.}(2018)\citenamefont {Bukov},
  \citenamefont {Day}, \citenamefont {Sels}, \citenamefont {Weinberg},
  \citenamefont {Polkovnikov},\ and\ \citenamefont
  {Mehta}}]{PhysRevX.8.031086}%
  \BibitemOpen
  \bibfield  {author} {\bibinfo {author} {\bibfnamefont {M.}~\bibnamefont
  {Bukov}}, \bibinfo {author} {\bibfnamefont {A.~G.~R.}\ \bibnamefont {Day}},
  \bibinfo {author} {\bibfnamefont {D.}~\bibnamefont {Sels}}, \bibinfo {author}
  {\bibfnamefont {P.}~\bibnamefont {Weinberg}}, \bibinfo {author}
  {\bibfnamefont {A.}~\bibnamefont {Polkovnikov}}, \ and\ \bibinfo {author}
  {\bibfnamefont {P.}~\bibnamefont {Mehta}},\ }\href {\doibase
  10.1103/PhysRevX.8.031086} {\bibfield  {journal} {\bibinfo  {journal} {Phys.
  Rev. X}\ }\textbf {\bibinfo {volume} {8}},\ \bibinfo {pages} {031086}
  (\bibinfo {year} {2018})}\BibitemShut {NoStop}%
\bibitem [{\citenamefont {Innocenti}\ \emph {et~al.}(2020)\citenamefont
  {Innocenti}, \citenamefont {Banchi}, \citenamefont {Ferraro}, \citenamefont
  {Bose},\ and\ \citenamefont {Paternostro}}]{Innocenti2020}%
  \BibitemOpen
  \bibfield  {author} {\bibinfo {author} {\bibfnamefont {L.}~\bibnamefont
  {Innocenti}}, \bibinfo {author} {\bibfnamefont {L.}~\bibnamefont {Banchi}},
  \bibinfo {author} {\bibfnamefont {A.}~\bibnamefont {Ferraro}}, \bibinfo
  {author} {\bibfnamefont {S.}~\bibnamefont {Bose}}, \ and\ \bibinfo {author}
  {\bibfnamefont {M.}~\bibnamefont {Paternostro}},\ }\href
  {https://iopscience.iop.org/article/10.1088/1367-2630/ab8aaf} {\bibfield
  {journal} {\bibinfo  {journal} {New J. Phys.}\ }\textbf {\bibinfo {volume}
  {22}},\ \bibinfo {pages} {065001} (\bibinfo {year} {2020})}\BibitemShut
  {NoStop}%
\bibitem [{\citenamefont {Harney}\ \emph {et~al.}(2018)\citenamefont {Harney},
  \citenamefont {Pirandola}, \citenamefont {Ferraro},\ and\ \citenamefont
  {Paternostro}}]{Harney2020}%
  \BibitemOpen
  \bibfield  {author} {\bibinfo {author} {\bibfnamefont {C.}~\bibnamefont
  {Harney}}, \bibinfo {author} {\bibfnamefont {S.}~\bibnamefont {Pirandola}},
  \bibinfo {author} {\bibfnamefont {A.}~\bibnamefont {Ferraro}}, \ and\
  \bibinfo {author} {\bibfnamefont {M.}~\bibnamefont {Paternostro}},\ }\href
  {https://iopscience.iop.org/article/10.1088/1367-2630/ab783d} {\bibfield
  {journal} {\bibinfo  {journal} {New J. Phys.}\ }\textbf {\bibinfo {volume}
  {22}},\ \bibinfo {pages} {045001} (\bibinfo {year} {2018})}\BibitemShut
  {NoStop}%
\bibitem [{\citenamefont {F\"osel}\ \emph {et~al.}(2018)\citenamefont
  {F\"osel}, \citenamefont {Tighineanu}, \citenamefont {Weiss},\ and\
  \citenamefont {Marquardt}}]{PhysRevX.8.031084}%
  \BibitemOpen
  \bibfield  {author} {\bibinfo {author} {\bibfnamefont {T.}~\bibnamefont
  {F\"osel}}, \bibinfo {author} {\bibfnamefont {P.}~\bibnamefont {Tighineanu}},
  \bibinfo {author} {\bibfnamefont {T.}~\bibnamefont {Weiss}}, \ and\ \bibinfo
  {author} {\bibfnamefont {F.}~\bibnamefont {Marquardt}},\ }\href {\doibase
  10.1103/PhysRevX.8.031084} {\bibfield  {journal} {\bibinfo  {journal} {Phys.
  Rev. X}\ }\textbf {\bibinfo {volume} {8}},\ \bibinfo {pages} {031084}
  (\bibinfo {year} {2018})}\BibitemShut {NoStop}%
\bibitem [{\citenamefont {Banchi}\ \emph {et~al.}(2018)\citenamefont {Banchi},
  \citenamefont {Grant}, \citenamefont {Rocchetto},\ and\ \citenamefont
  {Severini}}]{Banchi_2018}%
  \BibitemOpen
  \bibfield  {author} {\bibinfo {author} {\bibfnamefont {L.}~\bibnamefont
  {Banchi}}, \bibinfo {author} {\bibfnamefont {E.}~\bibnamefont {Grant}},
  \bibinfo {author} {\bibfnamefont {A.}~\bibnamefont {Rocchetto}}, \ and\
  \bibinfo {author} {\bibfnamefont {S.}~\bibnamefont {Severini}},\ }\href
  {\doibase 10.1088/1367-2630/aaf749} {\bibfield  {journal} {\bibinfo
  {journal} {New J. Phys.}\ }\textbf {\bibinfo {volume} {20}},\ \bibinfo
  {pages} {123030} (\bibinfo {year} {2018})}\BibitemShut {NoStop}%
\bibitem [{\citenamefont {Giordani}\ \emph {et~al.}(2019)\citenamefont
  {Giordani}, \citenamefont {Polino}, \citenamefont {Emiliani}, \citenamefont
  {Suprano}, \citenamefont {Innocenti}, \citenamefont {Majury}, \citenamefont
  {Marrucci}, \citenamefont {Paternostro}, \citenamefont {Ferraro},
  \citenamefont {Spagnolo},\ and\ \citenamefont {Sciarrino}}]{Giordani2019}%
  \BibitemOpen
  \bibfield  {author} {\bibinfo {author} {\bibfnamefont {T.}~\bibnamefont
  {Giordani}}, \bibinfo {author} {\bibfnamefont {E.}~\bibnamefont {Polino}},
  \bibinfo {author} {\bibfnamefont {S.}~\bibnamefont {Emiliani}}, \bibinfo
  {author} {\bibfnamefont {A.}~\bibnamefont {Suprano}}, \bibinfo {author}
  {\bibfnamefont {L.}~\bibnamefont {Innocenti}}, \bibinfo {author}
  {\bibfnamefont {H.}~\bibnamefont {Majury}}, \bibinfo {author} {\bibfnamefont
  {L.}~\bibnamefont {Marrucci}}, \bibinfo {author} {\bibfnamefont
  {M.}~\bibnamefont {Paternostro}}, \bibinfo {author} {\bibfnamefont
  {A.}~\bibnamefont {Ferraro}}, \bibinfo {author} {\bibfnamefont
  {N.}~\bibnamefont {Spagnolo}}, \ and\ \bibinfo {author} {\bibfnamefont
  {F.}~\bibnamefont {Sciarrino}},\ }\href
  {https://link.aps.org/doi/10.1103/PhysRevLett.122.020503} {\bibfield
  {journal} {\bibinfo  {journal} {Phys. Rev. Lett.}\ }\textbf {\bibinfo
  {volume} {122}},\ \bibinfo {pages} {020503} (\bibinfo {year}
  {2019})}\BibitemShut {NoStop}%
\bibitem [{\citenamefont {Giordani}\ \emph {et~al.}(2020)\citenamefont
  {Giordani}, \citenamefont {Suprano}, \citenamefont {Polino}, \citenamefont
  {Acanfora}, \citenamefont {Innocenti}, \citenamefont {Ferraro}, \citenamefont
  {Paternostro}, \citenamefont {Spagnolo},\ and\ \citenamefont
  {Sciarrino}}]{Giordani2020}%
  \BibitemOpen
  \bibfield  {author} {\bibinfo {author} {\bibfnamefont {T.}~\bibnamefont
  {Giordani}}, \bibinfo {author} {\bibfnamefont {A.}~\bibnamefont {Suprano}},
  \bibinfo {author} {\bibfnamefont {E.}~\bibnamefont {Polino}}, \bibinfo
  {author} {\bibfnamefont {F.}~\bibnamefont {Acanfora}}, \bibinfo {author}
  {\bibfnamefont {L.}~\bibnamefont {Innocenti}}, \bibinfo {author}
  {\bibfnamefont {A.}~\bibnamefont {Ferraro}}, \bibinfo {author} {\bibfnamefont
  {M.}~\bibnamefont {Paternostro}}, \bibinfo {author} {\bibfnamefont
  {N.}~\bibnamefont {Spagnolo}}, \ and\ \bibinfo {author} {\bibfnamefont
  {F.}~\bibnamefont {Sciarrino}},\ }\href {10.1103/PhysRevLett.124.160401}
  {\bibfield  {journal} {\bibinfo  {journal} {Phys. Rev. Lett.}\ }\textbf
  {\bibinfo {volume} {124}},\ \bibinfo {pages} {160401} (\bibinfo {year}
  {2020})}\BibitemShut {NoStop}%
\bibitem [{\citenamefont {der Plas}(2016)}]{book:rif.5}%
  \BibitemOpen
  \bibfield  {author} {\bibinfo {author} {\bibfnamefont {J.~V.}\ \bibnamefont
  {der Plas}},\ }\href {http://shop.oreilly.com/product/0636920034919.do}
  {\emph {\bibinfo {title} {Python Data Science Handbook}}}\ (\bibinfo
  {publisher} {O'Reilly Media},\ \bibinfo {year} {2016})\BibitemShut {NoStop}%
\bibitem [{\citenamefont {Sutton}\ and\ \citenamefont
  {Barto}(2015)}]{book:rif.7}%
  \BibitemOpen
  \bibfield  {author} {\bibinfo {author} {\bibfnamefont {R.~S.}\ \bibnamefont
  {Sutton}}\ and\ \bibinfo {author} {\bibfnamefont {A.~G.}\ \bibnamefont
  {Barto}},\ }\href@noop {} {\emph {\bibinfo {title} {Reinforcement Learning:
  An Introduction}}}\ (\bibinfo  {publisher} {The MIT Press Cambridge,
  Massachusetts, London, England},\ \bibinfo {year} {2015})\BibitemShut
  {NoStop}%
\bibitem [{\citenamefont {Dunjko}\ and\ \citenamefont
  {Briegel}(2018)}]{art:rif.21}%
  \BibitemOpen
  \bibfield  {author} {\bibinfo {author} {\bibfnamefont {V.}~\bibnamefont
  {Dunjko}}\ and\ \bibinfo {author} {\bibfnamefont {H.~J.}\ \bibnamefont
  {Briegel}},\ }\href {https://doi.org/10.1088/1361-6633/aab406} {\bibfield
  {journal} {\bibinfo  {journal} {Rep. Prog. Phys.}\ }\textbf {\bibinfo
  {volume} {81}},\ \bibinfo {pages} {074001} (\bibinfo {year}
  {2018})}\BibitemShut {NoStop}%
\bibitem [{\citenamefont {Marquardt}()}]{web:ANN}%
  \BibitemOpen
  \bibfield  {author} {\bibinfo {author} {\bibfnamefont {F.}~\bibnamefont
  {Marquardt}},\ }\href@noop {} {\enquote {\bibinfo {title} {Machine learning
  for physicists},}\ }\bibinfo {howpublished}
  {\url{"https://machine-learning-for-physicists.org"}},\ \bibinfo {note}
  {[Online; accessed April-2019]}\BibitemShut {NoStop}%
\bibitem [{\citenamefont {Jarzynski}(1997)}]{art:rif.2}%
  \BibitemOpen
  \bibfield  {author} {\bibinfo {author} {\bibfnamefont {C.}~\bibnamefont
  {Jarzynski}},\ }\href {http://link.aps.org/doi/10.1103/PhysRevLett.78.2690}
  {\bibfield  {journal} {\bibinfo  {journal} {Phys. Rev. Lett. {\bf 78}, 2690}\
  } (\bibinfo {year} {1997})}\BibitemShut {NoStop}%
\bibitem [{\citenamefont {Crooks}(1999)}]{art:rif.3}%
  \BibitemOpen
  \bibfield  {author} {\bibinfo {author} {\bibfnamefont {G.~E.}\ \bibnamefont
  {Crooks}},\ }\href {https://link.aps.org/doi/10.1103/PhysRevE.60.2721}
  {\bibfield  {journal} {\bibinfo  {journal} {Phys. Rev. E {\bf 60}, 2721}\ }
  (\bibinfo {year} {1999})}\BibitemShut {NoStop}%
\bibitem [{\citenamefont {Deffner}\ and\ \citenamefont
  {Lutz}(2011)}]{art:rif.5}%
  \BibitemOpen
  \bibfield  {author} {\bibinfo {author} {\bibfnamefont {S.}~\bibnamefont
  {Deffner}}\ and\ \bibinfo {author} {\bibfnamefont {E.}~\bibnamefont {Lutz}},\
  }\href {https://link.aps.org/doi/10.1103/PhysRevLett.107.140404} {\bibfield
  {journal} {\bibinfo  {journal} {Phys. Rev. Lett. {\bf 107}, 140404}\ }
  (\bibinfo {year} {2011})}\BibitemShut {NoStop}%
\bibitem [{\citenamefont {Deffner}\ and\ \citenamefont
  {Lutz}(2010)}]{art:rif.6}%
  \BibitemOpen
  \bibfield  {author} {\bibinfo {author} {\bibfnamefont {S.}~\bibnamefont
  {Deffner}}\ and\ \bibinfo {author} {\bibfnamefont {E.}~\bibnamefont {Lutz}},\
  }\href {https://link.aps.org/doi/10.1103/PhysRevLett.105.170402} {\bibfield
  {journal} {\bibinfo  {journal} {Phys. Rev. Lett. {\bf 105}, 170402}\ }
  (\bibinfo {year} {2010})}\BibitemShut {NoStop}%
\bibitem [{\citenamefont {Deffner}\ and\ \citenamefont
  {Campbell}(2019{\natexlab{b}})}]{art:rif.14}%
  \BibitemOpen
  \bibfield  {author} {\bibinfo {author} {\bibfnamefont {S.}~\bibnamefont
  {Deffner}}\ and\ \bibinfo {author} {\bibfnamefont {S.}~\bibnamefont
  {Campbell}},\ }\href {\doibase 10.1088/2053-2571/ab21c6} {\emph {\bibinfo
  {title} {Quantum Thermodynamics}}},\ 2053-2571\ (\bibinfo  {publisher}
  {Morgan and Claypool Publishers},\ \bibinfo {year} {2019})\BibitemShut
  {NoStop}%
\bibitem [{\citenamefont {Vedral}(2002)}]{Vedral2002}%
  \BibitemOpen
  \bibfield  {author} {\bibinfo {author} {\bibfnamefont {V.}~\bibnamefont
  {Vedral}},\ }\href {\doibase 10.1103/RevModPhys.74.197} {\bibfield  {journal}
  {\bibinfo  {journal} {Rev. Mod. Phys.}\ }\textbf {\bibinfo {volume} {74}},\
  \bibinfo {pages} {197} (\bibinfo {year} {2002})}\BibitemShut {NoStop}%
\bibitem [{\citenamefont {Nielsen}(2015)}]{book:rif.9}%
  \BibitemOpen
  \bibfield  {author} {\bibinfo {author} {\bibfnamefont {M.}~\bibnamefont
  {Nielsen}},\ }\href@noop {} {\emph {\bibinfo {title} {Neural Networks and
  Deep Learning}}}\ (\bibinfo  {publisher} {Determination Press},\ \bibinfo
  {year} {2015})\BibitemShut {NoStop}%
\bibitem [{\citenamefont {Talkner}\ \emph {et~al.}(2007)\citenamefont
  {Talkner}, \citenamefont {Lutz},\ and\ \citenamefont
  {H{\"a}nggi}}]{art:rif.7}%
  \BibitemOpen
  \bibfield  {author} {\bibinfo {author} {\bibfnamefont {P.}~\bibnamefont
  {Talkner}}, \bibinfo {author} {\bibfnamefont {E.}~\bibnamefont {Lutz}}, \
  and\ \bibinfo {author} {\bibfnamefont {P.}~\bibnamefont {H{\"a}nggi}},\
  }\href {https://link.aps.org/doi/10.1103/PhysRevE.75.050102} {\bibfield
  {journal} {\bibinfo  {journal} {Phys. Rev. E {\bf 75}, 050102(R)}\ } (\bibinfo
  {year} {2007})}\BibitemShut {NoStop}%
\bibitem{SM} {{S}upplemental {M}aterial reporting technical details and a comparison with other optimization methods is available from XXX.}
\bibitem [{\citenamefont {Mazzola}\ \emph {et~al.}(2013)\citenamefont
  {Mazzola}, \citenamefont {{De Chiara}},\ and\ \citenamefont
  {Paternostro}}]{art:rif.8}%
  \BibitemOpen
  \bibfield  {author} {\bibinfo {author} {\bibfnamefont {L.}~\bibnamefont
  {Mazzola}}, \bibinfo {author} {\bibfnamefont {G.}~\bibnamefont {{De
  Chiara}}}, \ and\ \bibinfo {author} {\bibfnamefont {M.}~\bibnamefont
  {Paternostro}},\ }\href
  {https://link.aps.org/doi/10.1103/PhysRevLett.110.230602} {\bibfield
  {journal} {\bibinfo  {journal} {Phys. Rev. Lett. {\bf 110}, 230602}\ }
  (\bibinfo {year} {2013})}\BibitemShut {NoStop}%
\bibitem [{\citenamefont {Chollet}\ \emph {et~al.}(2015)\citenamefont {Chollet}
  \emph {et~al.}}]{chollet2015keras}%
  \BibitemOpen
  \bibfield  {author} {\bibinfo {author} {\bibfnamefont {F.}~\bibnamefont
  {Chollet}} \emph {et~al.},\ }\href@noop {} {\enquote {\bibinfo {title}
  {Keras},}\ }\bibinfo {howpublished} {\url{https://keras.io}} (\bibinfo {year}
  {2015})\BibitemShut {NoStop}%
\bibitem [{\citenamefont {Kingma}\ and\ \citenamefont {Ba}(2014)}]{adam}%
  \BibitemOpen
  \bibfield  {author} {\bibinfo {author} {\bibfnamefont {D.}~\bibnamefont
  {Kingma}}\ and\ \bibinfo {author} {\bibfnamefont {J.}~\bibnamefont {Ba}},\
  }\href@noop {} {\bibfield  {journal} {\bibinfo  {journal} {International
  Conference on Learning Representations}\ } (\bibinfo {year}
  {2014})}\BibitemShut {NoStop}%
\bibitem [{\citenamefont {Zhang}\ \emph {et~al.}(2019)\citenamefont {Zhang},
  \citenamefont {Wei}, \citenamefont {Asad}, \citenamefont {Yang},\ and\
  \citenamefont {Wang}}]{Zhang_2019}%
  \BibitemOpen
  \bibfield  {author} {\bibinfo {author} {\bibfnamefont {X.-M.}\ \bibnamefont
  {Zhang}}, \bibinfo {author} {\bibfnamefont {Z.}~\bibnamefont {Wei}}, \bibinfo
  {author} {\bibfnamefont {R.}~\bibnamefont {Asad}}, \bibinfo {author}
  {\bibfnamefont {X.-C.}\ \bibnamefont {Yang}}, \ and\ \bibinfo {author}
  {\bibfnamefont {X.}~\bibnamefont {Wang}},\ }\href
  {http://dx.doi.org/10.1038/s41534-019-0201-8} {\bibfield  {journal} {\bibinfo
   {journal} {npj Quant. Inf.}\ }\textbf {\bibinfo {volume} {5}},\ \bibinfo
  {pages} {85} (\bibinfo {year} {2019})}\BibitemShut {NoStop}%
\bibitem [{\citenamefont {{Virtanen, {\it et al.}}}(2020)}]{2020SciPy-NMeth}%
  \BibitemOpen
  \bibfield  {author} {\bibinfo {author} {\bibfnamefont {P.}~\bibnamefont
  {{Virtanen, {\it et al.}}}},\ }\href {\doibase
  https://doi.org/10.1038/s41592-019-0686-2} {\bibfield  {journal} {\bibinfo
  {journal} {Nature Methods}\ }\textbf {\bibinfo {volume} {17}},\ \bibinfo
  {pages} {261} (\bibinfo {year} {2020})}\BibitemShut {NoStop}%
\end{thebibliography}
%
\onecolumngrid
\newpage
\appendix
\section*{Supplemental Material on \\
	"Reinforcement learning approach to non-equilibrium quantum thermodynamics"}
\subsection{Technical details and hyperparameters}
To facilitate reproducibility, we report here some technical details and the hyperparameters used in this work.
Starting with the networks structure, our Dense-layers Neural Network had 3 hidden layers of $100$ neurons and used a Rectified Linear Unit activation function, while our LSTM Neural Network consisted in $50$ LSTM units \cite{chollet2015keras} followed by a $30$ neurons dense layer with a Hyperbolic Tangent activation function. For both networks we set the activation function of the final layer to be a Hyperbolic Tangent, so that we could fix the maximum output value.

The training is performed using the "Adam" optimizer \cite{adam,chollet2015keras}.

For the first two approaches described in Section "Physical system and methodology", we put $\sigma=1$.  We further introduced a parameter $\epsilon$ such that, at each step, with a probability $\epsilon$, the agent takes a completely random action from a uniform distribution, ignoring the policy. This reduces the possibility for the agent to get stuck in a local maximum and leads to a better exploration of the space of possible actions. We fixed $\epsilon= 0.1$ up to the last set of 100 epochs, and then set it to zero.

When using the LSTM network we instead fixed $\sigma=0.1$ and $\epsilon=0$ and we subtracted a baseline to the reward $R \leftarrow R- \langle R \rangle_{batch}$, which usually leads to better convergence properties \cite{book:rif.7}.
\subsection{Comparison with other optimization techniques}
To further convince the reader of the validity of our methods, we report here a comparison with other optimization techniques. A systematic comparison of RL with different optimal control techniqes has been recently carried out in Ref. \cite{Zhang_2019} in the context of state preparation. In particular, the first two approaches proposed in Section "Physical system and methodology" can be seen as applications of the same RL methodology to a different physical problem. 

The third approach proposed is instead fundamentally different and hence a more in depth analysis is required.
Following a similar line of Ref. \cite{Zhang_2019}, we will consider the problem of state preparation. Here, however, we will consider the sistuation in which our knowledge of the system is minimal and no calculation nor numerical simulation is possible. The objective will be the preparation of a target state $|\psi_{target}(T)\rangle$ on a given time interval $[0,T]$, when the system starts from a state $|\psi(0)\rangle$.

We will use the same step-functions-like control described in Section "Physical system and methodology", extending Eq. \ref{control} to the case of multiple control terms
\begin{equation}\label{multicontrol}
H(t)=H_S+\sum_{j}f^{j}_\text{opt}(t)M^{j}_\text{opt}
\end{equation}
(Notice that here the Hamiltonian of the system is time-indipendent). As before, the optimization is performed over the values of $f^{j}_\text{opt}(t)$ on the intervals $[t_i, t_{i+1}[$. Our objective is the maximization of the function
\begin{equation}\label{max}
R=|\langle \psi_{target}(T)|\psi(T)\rangle|,
\end{equation}
or the minimization of the function $-R$.

In the following we will consider the situation in which a series of experiments are performed where \ref{max} can be evaluated at the end of the evolution and the optimization is performed accordingly. Our measure of performance will be the maximum $R$ achieved after a certain number of experiments.

We will compare the optimization performed with standard techniques \cite{2020SciPy-NMeth} where the cost function $-R$ is minimized numerically and the combination of RL and LSTM network proposed in this work where our agent aims to maximize the reward $R$. Our measure of "number of experiments" needed for the optimization will be the number of time a simulation of the system dynamics is performed for the maximization of $R$. We stress that this is not a measure of the time required for the numerical optimization but rather an indicator of how many experiments would be required in order to perform this optimization experimentally, without simulating the system dynamics.

The LSTM network structure will be the same used in the rest of this work (except for a bigger number of neurons in the output layer, determined by the number of control terms in Equation \ref{multicontrol}) and the only difference in the algorithm will be the following: instead of performing a final run over a batch of systems to find the optimal control functions we will take track of the maximum value of $R$ reached during the training and the corrisponding $f^j_{opt}(t_i)$, which will be considered to be our optimum.

\begin{figure}
	\includegraphics[width=0.5\columnwidth]{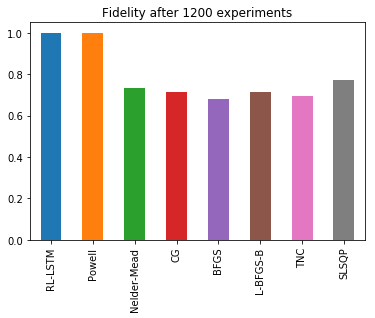}
	\caption{Results of the comparison between the RL-LSTM approach and different optimization algorithms \cite{2020SciPy-NMeth} for the single qubit state preparation problem described in the main text. It can be seen that the Fidelity (averaged over $10$ runs of the optimization) of the final state with the target state reached after $1200$ "experiments" is almost $1$ only when using the RL-LSTM technique or the "Powell Algorithm".}
	\label{comparison1}
\end{figure}

Our expectation is that the combination of RL and LSTM network, taking advantage of the nature of $\{f^j_{opt}(t_i)\}_i$ as a time series, can exploit causality and hence leads to a better optimization.
In Figure \ref{comparison1} and Table \ref{comparison2} we show the results of our simulations in the case of complete control ($\{M^j_{opt}\}_j$ are the Pauli Matrices and and the identity) for a qubit initially prepared in $|\psi(0)\rangle=|0\rangle$ (in the basis of $\sigma_z$) and a target state $|\psi_{target}(T)\rangle=\frac{1}{\sqrt{2}}(|1\rangle-|0\rangle)$.
Here the number of steps is $20$ and $\mu^{*}=1$. Similar constraints are applied to all the optimization algorithms considered. We can see that the approach proposed in this work can achieve larger values of fidelity in a lower number of experiments than any of the standard optimization techniques considered here.

\begin{figure}
	\includegraphics[width=0.5\columnwidth]{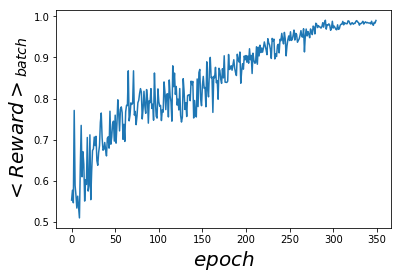}
	\caption{Learning curve of the agent for the problem of multiple state preparation on a qubit with complete control. For each experiment, the target state is chosen to be $|\psi_{target}(T)\rangle = a|0\rangle+\sqrt{1-a}|1\rangle$ with $a$ extracted from a uniform distribution in $[0,1]$ and the agent receives in input the value of $a$. Here the number of steps is $10$, $\mu^*=2$ and the training is carried out with a batch of $20$ systems. After $350$ epochs, the agent reaches an average value of the reward over the batch $\langle R \rangle_{batch}\approx0.990$.}
	\label{comparison3}
\end{figure}

Furthermore, a single agent can learn to prepare multiple target states on a single training phase, as shown in Figure \ref{comparison3}, while standard optimization techniques require a different optimizations for each state in which we want to prepare our system.
\begin{table}
	\begin{tabular}{ccc}
		\hline
		\hline
		number of experiments&  RL-LSTM & Powell \\
		\hline
		\hline
		$150$ & $0.905$ & $0.725$\\
		$200$ & $0.965$ & $0.776$\\
		$250$ & $0.986$ & $0.785$\\
		$300$ & $0.998$ & $0.902$\\
		\hline
	\end{tabular}
	\caption{Results for the single qubit state preparation problem (see the main text). Here we show the fidelity with the target state (averaged over $5$ runs of the optimization) for increasing number of experiments, reached using the RL-LSTM approach and the Powell algorithm (which, based on Figure \ref{comparison1} are the best approaches studied for this problem).}\label{comparison2}
\end{table}

Finally, it is worth looking at the effect on the performances of increasing system dimension. To this end, again following Ref. \cite{Zhang_2019}, we will consider the problem of excitation transfer between the ends of a qubits chain described by the Hamiltonian
\begin{equation}
H_S=\sum_{n=1}^{N}(\sigma^{n}_{x}\sigma^{n+1}_{x}+\sigma^{n}_{y}\sigma^{n+1}_{y}),
\end{equation}
where $\sigma^{n}_{j}$ is the $j-$Pauli matrix of the $n^{th}$ qubit and $N$ is the total number of qubits. We will appy a control
\begin{equation}
H_{opt}(t)=2\sum_{n}^{N}f^{n}_{opt}(t)\sigma^{n}_{z}.
\end{equation}
Regarding $\{f^j_{opt}(t_i)\}_i$ we will consider $20$ steps, $\mu^*=40$.
Results of the simulations are shown in Table \ref{comparison4}. It can be seen again that the approach proposed here outperforms the other existing techniques considered in this study.

\begin{table}
	\begin{tabular}{ccccc}
		\hline
		\hline
		N&  RL-LSTM & Powell & SLSQP & Nelder-Mead\\
		\hline
		\hline
		$2$ & $0.999$ & $0.891$& $0.790$ & $0.766$\\
		$3$ & $0.982$ & $0.825$& $0.568$ & $0.557$\\
		$4$ & $0.990$ & $0.633$& $0.271$ & $0.197$\\
		\hline
	\end{tabular}
	\caption{Fidelity with the target state (averaged over $5$ runs of the optimization) for the qubit chain problem (see the main text) for increasing number of qubits. Here we show the comparison between the most effective techniques studied.}\label{comparison4}
\end{table}
\end{document}